\documentclass[9pt,preprint]{sigplanconf}
\pdfoutput=1
\usepackage{balance}
\usepackage{amsmath}
\usepackage{amssymb}
\usepackage[mathscr]{euscript}
\usepackage{enumitem}
\usepackage{bussproofs}
\usepackage{mathrsfs}
\usepackage{wrapfig}

\usepackage{graphicx}
\usepackage{url}
\usepackage{mathpartir}
\usepackage{tikz}
\usetikzlibrary{shapes,backgrounds}

\usepackage{float}
\usepackage{wrapfig}
\usepackage{algorithm}
\usepackage{algorithmicx}     
\usepackage{algcompatible}    
\usepackage{listings}
\usepackage{time}
\usepackage{datetime}
\usepackage{amsmath}
\usepackage{amssymb}
\usepackage{lscape}
\usepackage{time}
\usepackage{datetime}
\usepackage{listings}

\usepackage{color}
\makeatletter 
\let\c@lofdepth\relax 
\let\c@lotdepth\relax 
\makeatother 
\usepackage[raggedright,small,sf,SF,hang]{subfigure}
\usepackage[toc,page,title,titletoc]{appendix}

\usepackage{float}
\usepackage{wrapfig}

\usepackage{extarrows}  

\usepackage{stmaryrd}   

\begin{document}



\renewcommand{\baselinestretch}{0.92} 
\newcommand{\clpr}{CLP(${\cal R}$)}
\newcommand{\np}{${\cal NP}$~}
\newcommand{\TR}{${\cal TR}$}
\newcommand{\CTR}{${\cal CTR}$}
\newcommand{\SR}{${\cal SR}$~}
\newcommand{\del}{{\em delta}}
\newcommand{\nil}{~{\rm nil}~}
\newcommand{\I}{{\cal I}~}
\newcommand{\exedout}{%
  \rule{0.8\textwidth}{0.5\textwidth}%
}

\newcommand{\ignore}[1]{}

\newcommand{\stuff}[1]{
        \begin{minipage}{6in}
        {\tt \samepage
        \begin{tabbing}
        \hspace{3mm} \= \hspace{3mm} \= \hspace{3mm} \= \hspace{3mm} \= \hspace{3mm} \= \hspace{3mm} \=
\hspace{3mm} \= \hspace{3mm} \= \hspace{3mm} \= \hspace{3mm} \= \hspace{3mm} \= \hspace{3mm} \= \hspace{3mm} \= \kill
        #1
        \end{tabbing}
       }
        \end{minipage}
}

\newcommand{\mystuff}[1]{
        \begin{minipage}[b]{6in}
        {\tt \samepage
        \begin{tabbing}
        \hspace{3mm} \= \hspace{3mm} \= \hspace{3mm} \= \hspace{3mm} \= \hspace{3mm} \= \hspace{3mm} \=
\hspace{3mm} \= \hspace{3mm} \= \hspace{3mm} \= \hspace{3mm} \= \hspace{3mm} \= \hspace{3mm} \= \hspace{3mm} \= \kill
        #1
        \end{tabbing}
       }
        \end{minipage}
}

\newcommand{\newmystuff}[1]{
        \begin{minipage}[b]{6in}
        {\tt \samepage
        \begin{tabbing}
        \hspace{3mm} \= \hspace{3mm} \= \hspace{3mm} \= \hspace{3mm} \= \hspace{3mm} \= \hspace{3mm} \=
\hspace{3mm} \= \hspace{3mm} \= \hspace{3mm} \= \hspace{3mm} \= \hspace{3mm} \= \hspace{3mm} \= \hspace{3mm} \= \kill
        #1
        \end{tabbing}
       }
        \end{minipage}
}

\newcommand{\Rule}[2]{\genfrac{}{}{0.5pt}{}
{{\setlength{\fboxrule}{0pt}\setlength{\fboxsep}{3mm}\fbox{$#1$}}}
{{\setlength{\fboxrule}{0pt}\setlength{\fboxsep}{3mm}\fbox{$#2$}}}}

\newcommand{\rct}{\mbox{\it clp}}
\newcommand{\lif}{\ {\tt:\!\!-} \ }
\newcommand{\ctbl}{\,{\mbox{\footnotesize$|$}\tt-\!\!\!\!>}\,}
\newlength{\colwidth}
\setlength{\colwidth}{.47\textwidth}


\newcommand{\eqdef}{\ensuremath{\overset{\textit{def}}{=}}}
\newcommand{\sys}{\mbox{\bf sys}}
\newcommand{\var}{\mbox{\frenchspacing \it var}}
\newcommand{\lfp}{\mbox{\it \frenchspacing lfp}}
\newcommand{\gfp}{\mbox{\it \frenchspacing gfp}}
\newcommand{\assert}{\mbox{\it \frenchspacing assert}}
\newcommand{\buffer}{\mbox{\it \frenchspacing Buffer}}
\newcommand{\clpif}{\mbox{\tt :-}}


\newtheorem{definition}{Definition}
\newtheorem{theorem}{Theorem}
\newtheorem{lemma}{Lemma}
\newtheorem{proposition}{Proposition}
\newtheorem{corollary}{Corollary}
\newtheorem{example}{Example}
\newtheorem{myproof}{Proof Outline}

\newcommand{\QED}{\nolinebreak\hskip 1em
        \framebox[0.5em]{\rule{0ex}{1.0ex}}}
\newcommand{\qed}{\nolinebreak\hskip 1em
        \framebox[0.5em]{\rule{0ex}{1.0ex}}}

\newcommand{\close}{{\frenchspacing close }}
\newcommand{\rules}{{\frenchspacing rules }}
\newcommand{\lhs}{{\frenchspacing lhs }}
\newcommand{\rhs}{{\frenchspacing rhs }}
\newcommand{\wrt}{{\frenchspacing wrt. }}
\newcommand{\fold}{\mbox{\it\frenchspacing old}}
\newcommand{\exactunfold}{\mbox{\it\frenchspacing exactunfold}}

\newcommand{\mgu}{{\frenchspacing mgu }}
\newcommand{\size}{{\frenchspacing size }}
\newcommand{\obs}{{\frenchspacing obs }}
\newcommand{\trace}{{\frenchspacing trace }}

\newcommand{\xxx}{\mbox{\Large $\vartriangleright$}}

\newcommand{\undeniable}{\mbox{\Large $\vartriangleright$}\hspace{-8pt}\raisebox{2pt}{\tiny u}~~}
\newcommand{\inevitable}{\mbox{\Large $\vartriangleright$}\hspace{-8pt}\raisebox{2pt}{\tiny i}~~\,}

\newcommand{\reachable}{\mbox{\Large $\vartriangleright$}\hspace{-9pt}\raisebox{1pt}{\tiny *}~~\,}

\newcommand{\lbr}{\mbox{$[\![$}}
\newcommand{\rbr}{\mbox{$]\!]$}}
\newcommand{\ct}[1]{\mbox{\lbr$\vec{#1}$\rbr}}
\newcommand{\cs}[1]{\mbox{\lbr#1\rbr}}
\newcommand{\clp}[1]{{\it \frenchspacing clp}(\mbox{$#1$})}
\newcommand{\pp}[1]{\mbox{{\color{blue}{$\langle$#1$\rangle$}}}}

\newcommand{\ao}{\mbox{${\cal A}$}}
\newcommand{\co}{\mbox{${\cal C}$}}
\newcommand{\po}{\mbox{${\cal P}$}}

\newcommand{\tA}{{\tilde{A}}}
\newcommand{\tB}{{\tilde{B}}}
\newcommand{\tC}{{\tilde{C}}}

\newcommand{\A}{\mbox{$\cal A$}}
\newcommand{\B}{\mbox{$\cal B$}}
\newcommand{\D}{\mbox{$\cal D$}}
\newcommand{\E}{\mbox{$\cal E$}}
\newcommand{\M}{\mbox{$\cal M$}}
\newcommand{\V}{\mbox{$\cal V$}}
\newcommand{\unfold}{\mbox{\sc unfold}}

\newcommand{\G}{\mbox{$\cal G$}}
\newcommand{\Gone}{\mbox{${\cal G}_{\!1}$}}
\newcommand{\Gi}{\mbox{${\cal G}_{\!i}$}}
\newcommand{\Gn}{\mbox{${\cal G}_{\!n}$}}

\newcommand{\GL}{\mbox{${\cal G}_{\!L}$}}
\newcommand{\GR}{\mbox{${\cal G}_{\!R}$}}

\newcommand{\HH}{\mbox{$\cal H$}}
\newcommand{\HHone}{\mbox{${\cal H}_{\!1}$}}
\newcommand{\HHi}{\mbox{${\cal H}_{\!i}$}}
\newcommand{\HHj}{\mbox{${\cal H}_{\!j}$}}
\newcommand{\HHn}{\mbox{${\cal H}_{\!n}$}}
\newcommand{\HHm}{\mbox{${\cal H}_{\!m}$}}

\newcommand{\Left}{\mbox{${\cal L}$}}
\newcommand{\Right}{\mbox{${\cal R}$}}

\newcommand{\LeftD}{\mbox{${p}$}}
\newcommand{\RightD}{\mbox{${q}$}}

\newcommand{\Mone}{\mbox{${\cal M}_{\!1}$}}
\newcommand{\Mtwo}{\mbox{${\cal M}_{\!2}$}}
\newcommand{\MMi}{\mbox{${\cal M}_{\!i}$}}
\newcommand{\MMn}{\mbox{${\cal M}_{\!n}$}}
\newcommand{\MMnone}{\mbox{${\cal M}_{\!n-1}$}}

\newcommand{\PsiL}{\mbox{$\Psi_{\!L}$}}
\newcommand{\PsiR}{\mbox{$\Psi_{\!R}$}}
\newcommand{\PsiB}{\mbox{$\Psi_{\!B}$}}

\newcommand{\Af}{A_{\!f}}
\newcommand{\Ax}{A_{\!1}}
\newcommand{\Nf}{N_{\!f}}
\newcommand{\Xf}{X_{\!f}}
\newcommand{\Yf}{Y_{\!f}}
\newcommand{\Hxxx}{H_{\!f}}

\newcommand{\cts}{\mbox{$\mapsto$}}
\newcommand{\sepimp}{\mbox{$-\!*$}}

\newcommand{\arr}[3]{\mbox{$\langle \mbox{#1,#2,#3} \rangle$}}
\newcommand{\triple}[3]{\langle{#1},{#2},{#3}\rangle}
\newcommand{\aquadruple}[4]{\langle{#1},{#2},{#3},{#4}\rangle}
\newcommand{\baeq}[2]{\mbox{$=_{\[#1 .. #2\]}$}}

\newcommand{\ite}[3]{\mbox{\textit{ite}}({#1},{#2},{#3})}
\newcommand{\pred}[1]{\mbox{\textit{#1}}}
\newcommand{\ptab}{~~~~}

\newlength{\vitelen}
\settowidth{\vitelen}{\mbox{\textit{ite}}(}
\newcommand{\vite}[3]{\begin{array}[t]{l}\mbox{\textit{ite}}({#1},\\
\hspace{\vitelen}{#2},\\
\hspace{\vitelen}{#3}
    \end{array}}

\newcommand{\Hfx}{H_{\!f}}
\newcommand{\Hx}{H_{\!1}}
\newcommand{\Hxx}{H_{\!2}}
\newcommand{\Pf}{P_{\!f}}
\newcommand{\Pz}{P_{\!0}}
\newcommand{\Px}{P_{\!1}}

\newcommand{\Ix}{I_{\!1}}
\newcommand{\Ixx}{I_{\!2}}
\newcommand{\Jf}{J_{\!f}}
\newcommand{\Jx}{J_{\!1}}
\newcommand{\Jxx}{J_{\!2}}

\newcommand{\witness}{\mbox{$\omega$}}
\newcommand{\transformer}{\mbox{$\Delta$}}

\newcommand{\func}[1]{\mbox{\textsf{#1}}}

\floatstyle{boxed}
\restylefloat{figure}

%

\renewcommand\labelitemi{$\bullet$}
\renewcommand\labelitemii{\normalfont\bfseries --}

\renewcommand\floatpagefraction{.9}
\renewcommand\topfraction{.9}
\renewcommand\bottomfraction{.9}
\renewcommand\textfraction{.1}   
\setcounter{totalnumber}{50}
\setcounter{topnumber}{50}
\setcounter{bottomnumber}{50}

\raggedbottom

\makeatletter
\newcommand{\manuallabel}[2]{\def\@currentlabel{#2}\label{#1}}
\makeatother

\newcounter{chapcount}
\newcommand{\chapcountreset}{\setcounter{chapcount}{0}}
\chapcountreset

\newcounter{excount}
\newcommand{\exreset}{\setcounter{excount}{0}}
\exreset
\newcommand{\newexample}[1]{\addtocounter{excount}{1}
{\vspace{2mm} \noindent \mbox{{\scriptsize EXAMPLE} \examplecount~\emph{#1}:}}}

\newcommand{\examplecount}{\arabic{excount}}

\newcommand{\exlabel}[1]{\manuallabel{#1}{\examplecount}}

\newcommand{\todo}[1]{
\noindent \framebox{\textbf{ #1}}
\newline
}

\newcommand{\algorithmicinput}{\textbf{Input:~}}
\newcommand{\algorithmicoutput}{\textbf{Output:~}}
\newcommand{\algorithmicglobal}{\textbf{Globally:~}}
\newcommand{\algorithmicfunction}{\textbf{Function}\ }
\newcommand{\algorithmicfunctionend}{\textbf{EndFunction}\ }
\newcommand{\memoed}{\mbox{\func{memoed}}}
\newcommand{\memoize}{\mbox{\func{memoize}}}

\newcommand{\memotable}{\mbox{\pred{Table}}}
\newcommand{\wpc}{\mbox{$\func{pre}$}}
\newcommand{\pre}[2]{\mbox{$\func{pre}(#1, #2)$}}

\newcommand{\interp}{\mbox{$\pred{Intp}$}}
\newcommand{\emanate}{\mbox{\func{outgoing}}}
\newcommand{\absiteration}{\mbox{\func{loop\_end}}}
\newcommand{\step}{\mbox{\func{TransStep}}}
\newcommand{\compress}{\mbox{\func{JoinVertical}}}
\newcommand{\join}{\mbox{\func{JoinHorizontal}}}

\newcommand{\program}{\mbox{$\cal P$}}
\newcommand{\N}{\mbox{$\cal N$}}

\newcommand{\Assign}{\mbox{:=}}
\newcommand{\pair}[2]{\langle{#1};{#2}\rangle}
\newcommand{\tuple}[3]{\langle{#1},{#2},{#3}\rangle}
\newcommand{\quadruple}[4]{[{#1},{#2},{#3},{#4}]}
\newcommand{\fivetuple}[5]{[{#1},{#2},{#3},{#4},{#5}]}

\newcommand{\afivetuple}[5]{\langle{#1},{#2},{#3},{#4},{#5}\rangle}

\newcounter{pppcount}
\newcommand{\pppreset}{\setcounter{pppcount}{0}}
\newcommand{\ppp}{\refstepcounter{pppcount}\mbox{{\color{blue}$\langle\arabic{pppcount}\rangle$}}}

\newcommand{\note}[1]{\marginpar{\color{red}{#1}}}
\newcommand{\tnote}[1]{\marginpar{\color{green}{#1}}}

\newcommand{\resource}{\mbox{\textsf r}}
\newcommand{\timing}{\mbox{\textsf t}}

\newcommand{\summarize}{\mbox{\cal S}}

\newcommand{\For}{\mbox{\textsf{for}}}
\newcommand{\Else}{\mbox{\textsf{else}}}
\newcommand{\If}{\mbox{\textsf{if}}}

\newcommand{\myiff}{\mbox{\texttt{iff}}}

\newcommand{\void}{\mbox{\textsf{void}}}
\newcommand{\assume}[1]{\mbox{\textsf{assume(#1)}}}
\newcommand{\assign}[2]{\mbox{\textsf{#1 := #2}}}

\newcommand{\exec}{\mbox{\textsf{exec}}}

\newcommand{\transition}[3]{#1~\xlongrightarrow[]{#3}~#2}
\newcommand{\shorttransition}[2]{#1~\xrightarrow[]{}~#2}
\newcommand{\trans}{\longrightarrow}
\newcommand{\shorttrans}{\rightarrow}
\newcommand{\translabel}[1]{\xlongrightarrow[]{#1}}
\newcommand{\loc}{\mbox{$\ell$}}
\newcommand{\locations}{\mbox{${\cal L}$}}

\newcommand{\transsystem}{\mbox{$\mathcal{P}$}}
\newcommand{\newtranssystem}{\mbox{$\mathcal{G}$}}

\newcommand{\symstate}{\mbox{$s$}}
\newcommand{\pci}[1]{\mbox{$\loc_{#1}$}}
\newcommand{\pc}{\mbox{\loc}}
\newcommand{\pcend}{\mbox{$\loc_{\textsf{end}}$}}
\newcommand{\pcerror}{\mbox{$\loc_{\textsf{error}}$}}
\newcommand{\pcstart}{\mbox{$\loc_{\textsf{start}}$}}
\newcommand{\pathcond}{\mbox{$\Pi$}}
\newcommand{\store}{\mbox{$\sigma$}}
\newcommand{\pathcondbar}{\mbox{$\overline{\Pi}$}}
\newcommand{\storebar}{\mbox{$\overline{h}$}}
\newcommand{\symstatebar}{\mbox{$\overline{\symstate}$}}
\newcommand{\mapstatetoformula}[1]{\mbox{$\llbracket {#1} \rrbracket$}}

\renewcommand{\path}{\mbox{$\theta$}}

\newcommand{\typevar}{\mbox{\emph{Vars}}}
\newcommand{\typesymvar}{\mbox{\emph{SymVars}}}
\newcommand{\typeop}{\mbox{\emph{Ops}}}
\newcommand{\typefo}{\mbox{\emph{FO}}}
\newcommand{\typeterms}{\mbox{\emph{Terms}}}
\newcommand{\typestate}{\mbox{\emph{States}}}
\newcommand{\typesymbstate}{\mbox{\emph{SymStates}}}
\newcommand{\typesympath}{\mbox{\emph{SymPaths}}}
\newcommand{\true}{\mbox{\frenchspacing \it true}}
\newcommand{\false}{\mbox{\frenchspacing \it false}}
\newcommand{\typebool}{\mbox{\emph{Bool}}}
\newcommand{\typeint}{\mbox{\emph{Int}}}
\newcommand{\typenat}{\mbox{\emph{Nat}}}
\newcommand{\typekeys}{\mbox{$\mathcal{K}$}}
\newcommand{\typevoid}{\mbox{\emph{Void}}}

\newcommand{\eval}[2]{\llbracket {#1} \rrbracket_{#2}}
\newcommand{\define}{\mbox{~$\triangleq$~}}
\newcommand{\unknown}{\mbox{\textsf{$\cdot$}}}

\newcommand{\Intpsymbol}{\mbox{$\overline{\Psi}$}}
\newcommand{\InvariantFunc}{\mbox{\textsf{invariant}}}
\newcommand{\InvariantSym}{\mbox{$\mathcal{I}$}}
\newcommand{\ConflictSym}{\mbox{$\mathcal{C}$}}
\newcommand{\ContextSym}{\mbox{$\mathcal{O}$}}
\newcommand{\modifies}{\mbox{\textsf{\textsc{Modifies}}}}
\newcommand{\havoc}{\mbox{\textsf{\textsc{Havoc}}}}
\newcommand{\getvars}{\mbox{\textsf{var}}}

\newcommand{\state}[1]{\mbox{{\small \textsf{#1}}}}
\newcommand{\safety}{\mbox{$\psi$}}
\newcommand{\cons}{\mbox{$\phi$}}
\newcommand{\target}{\mbox{$\gamma$}}
\newcommand{\solutions}{\mbox{$\Gamma$}}
\newcommand{\id}[2]{\mbox{\textsf{Id}($#1, #2$)}}

\renewcommand{\If}{\mbox{\textbf{if}}}
\newcommand{\Endif}{\mbox{\textbf{endif}}}
\newcommand{\Return}{\mbox{\textbf{return}}}
\newcommand{\Then}{\mbox{\textbf{then}}}
\renewcommand{\Else}{\mbox{\textbf{else}}}
\newcommand{\Foreach}{\mbox{\textbf{foreach}}}
\newcommand{\While}{\mbox{\textbf{while}}}
\newcommand{\Do}{\mbox{\textbf{do}}}
\newcommand{\Endfor}{\mbox{\textbf{endfor}}}
\newcommand{\Endwhile}{\mbox{\textbf{endwhile}}}

\newcommand{\ptr}[2]{#1$\!\rightarrow\!\,$#2}
\newcommand{\ptrmath}[2]{#1\!\rightarrow\!#2}

\newcommand{\sat}{{\textsc{sat}}}
\newcommand{\unsat}{\textsc{unsat}}
\newcommand{\smt}{{\textsc{smt}}}
\newcommand{\por}{{\textsc{por}}}
\newcommand{\dpor}{{\textsc{dpor}}} 
\newcommand{\si}{{\textsc{si}}}
\newcommand{\ti}{{\textsc{ti}}}
\newcommand{\cegar}{{\textsc{cegar}}}
\newcommand{\al}{{\textsc{al}}}
\newcommand{\rcsp}{{\textsc{rcsp}}}
\newcommand{\wcet}{{\textsc{wcet}}}
\newcommand{\ilp}{{\textsc{ilp}}}
\newcommand{\ipet}{{\textsc{ipet}}}
\newcommand{\cfg}{{\textsc{cfg}}}
\newcommand{\saturn}{{\textsc{saturn}}}
\renewcommand{\dag}{{\textsc{dag}}}

\newcommand{\dryad}{{\textsc{dryad}}}
\newcommand{\ai}{{\textsc{ai}}}

\renewcommand{\lhs}{{\small \textsf{LHS}}}
\renewcommand{\rhs}{{\small \textsf{RHS}}}

\newcommand{\nextpointer}{{\sf {nxt}}}
\newcommand{\lseg}{{\sf {ls}}}
\newcommand{\listn}{{\sf {list}}}
\newcommand{\lslast}{{\sf {list\_last}}}
\newcommand{\lsapp}{{\sf {list\_append}}}
\newcommand{\lsegleft}{${\sf \widehat{ls}}$}
\newcommand{\lsegleftmath}{{\sf \widehat{ls}}}

\newcommand{\sortedls}{${\sf {sorted\_ls}}$}
\newcommand{\sortedlist}{${\sf {sorted\_list}}$}

\newcommand{\dlseg}{{\sf {dls}}}
\newcommand{\dlistn}{{\sf {dlist}}}
\newcommand{\dlsegleft}{${\sf \widehat{dls}}$}

\newcommand{\masrt}[0]{\models_\mathrm{asrt}}
\newcommand{\mclph}[0]{\models}
\newcommand{\dom}[0]{\mathit{dom}}
\newcommand{\range}[0]{\mathit{range}}
\newcommand{\vars}[0]{\mathit{vars}}
\newcommand{\spc}[0]{\mathit{sp}}

\newcommand{\lm}[0]{\mathit{lm}}

\newcommand{\HT}[0]{\HH}
\newcommand{\R}[0]{\mathbb{R}}
\newcommand{\Z}[0]{\mathbb{Z}}
\newcommand{\MM}[0]{\mathcal{M}}

\newcommand{\syn}[1]{\langle \textit{#1} \rangle}
\newcommand{\Vars}{\set{ProgVars}}
\newcommand{\set}[1]{\mathsf{#1}}
\newcommand{\hsolve}[0]{\mathsf{hsolve}}

\newcommand{\GV}[0]{\mathsf{GV}}
\newcommand{\cp}[1]{\mathtt{#1}}
\newcommand{\heap}[0]{\bar{\mathcal{H}}}
\newcommand{\old}[0]{\mathbf{old}}
\newcommand{\init}[0]{\mathbf{i}}
\newcommand{\one}[2]{(#1 \mapsto #2)}
\newcommand{\interpret}[0]{\mathcal{I}}
\newcommand{\States}[0]{\mathsf{States}}
\newcommand{\Stores}[0]{\mathsf{Stores}}
\newcommand{\Heaps}[0]{\mathsf{Heaps}}
\newcommand{\Exprs}[0]{\mathsf{Exprs}}
\newcommand{\config}[2]{\langle #1, #2 \rangle}
\newcommand{\striple}[3]{\{#1\}#2\{#3\}}
\newcommand{\btriple}[3]{\big\{#1\big\}~#2~\big\{#3\big\}}
\newcommand{\vtriple}[3]{\begin{array}{c}\{#1\}\\#2\\\{#3\}\end{array}}
\newcommand{\vvtriple}[3]{\begin{array}{c}\{#1\}\\#2\\\left\{#3\right\}\end{array}}
\newcommand{\clpdot}[0]{\texttt{.}}
\newcommand{\prop}[0]{\Longrightarrow}
\newcommand{\simp}[0]{\Longleftrightarrow}
\newcommand{\hin}[3]{\mathsf{in}(#1, #2, #3)}

\newcommand{\hemp}[0]{\Omega}
\newcommand{\heq}[0]{\bumpeq}
\newcommand{\hsep}[0]{{\symbol{42}}}

\newcommand{\red}{\sc reduct}

\def\parmap{\rightharpoonup}

\floatsep 2mm plus 1mm minus 0mm
\dblfloatsep 6pt plus 2pt minus 2pt 
\textfloatsep 2mm plus 1mm minus 0mm
\dbltextfloatsep 6pt plus 2pt minus 2pt
\setlength{\intextsep}{4pt plus 2pt minus 0pt}
\belowcaptionskip 4pt plus 2pt minus 0pt
\abovecaptionskip 4pt plus 2pt minus 0pt




\title{Automating Proofs of Data-Structure Properties\\{in Imperative Programs}}

\authorinfo{Duc-Hiep Chu}
           {National University of Singapore}
           {hiepcd@comp.nus.edu.sg}
\authorinfo{Joxan Jaffar}
           {National University of Singapore}           
           {joxan@comp.nus.edu.sg} 
\authorinfo{Minh-Thai Trinh}
           {National University of Singapore}
           {trinhmt@comp.nus.edu.sg}

\maketitle




\begin{abstract}

We consider the problem of automated
reasoning about dynamically manipulated data structures.
The state-of-the-art methods are limited to the unfold-and-match (U+M) paradigm,
where predicates  are transformed via (un)folding operations induced from 
their definitions before being treated as \emph{uninterpreted}.
However, proof obligations from verifying programs with iterative loops 
and multiple function calls often do not succumb to this paradigm.
Our contribution is a proof method which -- beyond U+M -- 
performs \emph{automatic} formula re-writing   
by treating previously encountered obligations in each proof path 
as possible \emph{induction hypotheses}.
This enables us, for the first time, to systematically  
reason about a wide range of obligations, arising from practical 
program verification.

We demonstrate the power of our proof rules on 
commonly used \emph{lemmas}, thereby close the remaining 
gaps in existing state-of-the-art systems. Another impact, 
probably more important, is that our method regains the power 
of compositional reasoning, and shows that the usage of user-provided 
lemmas is no longer needed for the existing set of benchmarks. This not only
removes the burden of coming up with the appropriate lemmas, but also significantly 
boosts up the verification process, since lemma applications, coupled with unfolding, 
often induce very large search space.

%

\ignore{
In this paper, we first argue that in order to support general \emph{user-defined} predicates, 
a proof framework must be able to relate between \emph{different} predicates.
The state-of-the-art, classified under \emph{unfold-and-match} (U+M) paradigm, are
severely limited in this regard. As a matter of fact, \emph{unjustified} use of abstractions
-- usually in the form of lemmas/axioms -- is employed, making these existing systems
no longer fully automated.
  
Our main contribution then is an algorithm, which subsumes (U+M), while
overcomes the main limitation of state-of-the-arts by automatically employing induction proof steps
with respect to dynamically generated induction hypotheses.  Though
necessarily still incomplete, the algorithm provides a new level of automation.
}

\ignore{
  We consider the problem of automated program verification with
  emphasis on reasoning about dynamically manipulated data structures.
  We begin with an existing specification language which has two key
  features: (a) the use of explicit heap variables, and (b) user
  defined recursive properties in a wrapper logic language.  The
  language provides a new-level of expressiveness for specifying
  properties of heap manipulations.

The main contribution, however, is an algorithm to automatically prove
verification conditions when formulas in the specification language
are used as assertions in programs.  More precisely, we consider
verification of heap manipulating programs that combines user written
modular contracts and loop invariants 
with completely automated theorem proving
of the resulting verification conditions.  The key feature of the
algorithm is the ability to automatically employ induction proof steps
with respect to dynamically generated induction hypotheses.  Though
necessarily still incomplete, the algorithm provides a new level of
automation.
}

\end{abstract}



\section{Introduction}
\label{sec:intro}

We consider the 
automated verification of imperative programs 
with emphasis on reasoning about the functional correctness of 
dynamically manipulated data structures.
In this problem domain, pre/post conditions are specified 
for each function and an invariant is given for each loop
before the reasoning system automatically checks if the program code 
is correct wrt. the given annotations.
The dynamically modified heap 
poses a big challenge for logical methods. 
This is because typical correctness properties often
require complex combinations of structure, data, and \emph{separation}.

%


Automated proofs of data structure properties  
--- usually formalized using Separation Logic (or the alike) and extended with 
\emph{user-defined} recursive predicates 
--- ``rely on decidable sub-classes together with the corresponding proof systems 
based on (un)folding strategies for recursive definitions'' \cite{navarro11pldi}. 
Informally, in the regard of handling recursive predicates,
the state-of-the-art~\cite{nguyen10shape3,madhusudan12dryad,qiu13dryad,wies13cav,pek14pldi},
collectively called unfold-and-match (U+M) paradigm,
employ the basic but systematic transformation steps 
of \emph{folding} and \emph{unfolding} the rules.

A proof, using U+M, succeeds when we find 
successive applications of these transformation steps 
that produce a final formula which is \emph{obviously} provable.  
This usually means that either (1) there is no recursive predicate in the 
RHS of the proof obligation and a direct proof can be achieved by
consulting some generic \smt{} solver; 
or (2) no special consideration is needed on any occurrence of a predicate
appearing in the final formula.  For example, if
${\tt p}(\tilde{u}) ~ \wedge \cdots \models {\tt p}(\tilde{v})$
is the formula, then this is obviously provable if $\tilde{u}$ and $\tilde{v}$ were
\emph{unifiable} (under an appropriate theory governing the meaning of
the expressions $\tilde{u}$ and $\tilde{v}$).
In other words, we have performed 
``formula abstraction''~\cite{madhusudan12dryad} by treating
the recursively defined term ${\tt p}()$ as \emph{uninterpreted}. 


%

\subsection*{Proving Relationship between Unmatchable Predicates}


We say, informally, a proof obligation involves \emph{unmatchable} predicates
if there exists a recursively defined predicate in the RHS which cannot be transformed,
via folding/unfolding, to one that is unifiable with some predicate in the LHS.
It can be seen that, U+M (folding/unfolding together with formula abstraction)
\emph{cannot} prove relationship between unmatchable predicates.

Let us now highlight scenarios, which are \emph{ubiquitous} in
realistic programs, and often lead to proof obligations 
involving unmatchable predicates.
We first articulate them briefly, and then proceed with more examples.

\begin{description}
\item[\textit{Recursion Divergence}]
\ \\
when the ``recursion'' in the recursive rules
is structurally dissimilar to the program code.
\item[\textit{Generalization of Predicate}]
\ \\
when the predicate describing a loop invariant or
a function needs to be used later to prove
a weaker property.
\end{description}

\noindent
First consider ``recursion divergence''.
This happens often with \emph{iterative} programs and when the 
predicates are not \emph{unary}, i.e., they relate two or 
more pointer variables, from which the program code traverse or 
manipulate the data structure in directions 
different from the definition.

\begin{figure}
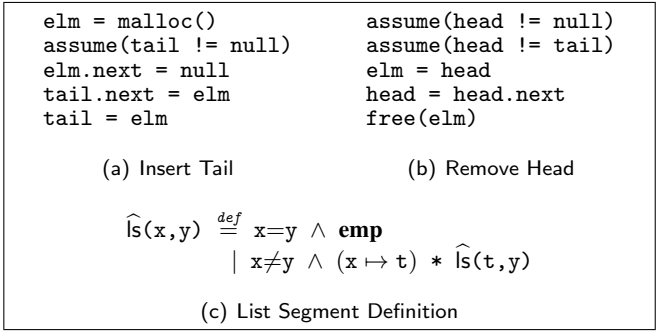

\centering
\subfigure[Insert Tail] {
\mystuff
{

elm = malloc() \\
assume(tail != null) \\
elm.next = null \\
tail.next = elm\\
tail = elm\\
}
\label{fig:intro:queue1}
}
\hspace{5mm}
\subfigure[Remove Head] {
\mystuff
{
assume(head != null) \\
assume(head != tail) \\
elm = head \\
head = head.next \\
free(elm)\\
}
\label{fig:intro:queue2}
} \\
\centering
\subfigure[List Segment Definition]{
\mystuff{
\lsegleft(x,y) $\eqdef$  x$=$y $\wedge$ {\bf emp}\\
\> \> \>        $|$        x$\neq$y $\wedge$ $\one{\tt x}{\tt t}$ \hsep ~\lsegleft(t,y) \\
}
\label{fig:intro:def}
}

\vspace{-2mm}
\caption{Implementation of a Queue}
\label{fig:intro:queue}
\end{figure}

To illustrate, Fig.~\ref{fig:intro:queue} shows the implementation of a queue using list segments, 
extracted from the open source program {\tt OpenBSD/queue.h}.
Two operations of interest: (1) adding a new element into
the end of a non-empty queue ({\tt enqueue}, Fig.~\ref{fig:intro:queue1}); 
(2) deleting an element at the beginning of a non-empty queue ({\tt dequeue}, Fig.~\ref{fig:intro:queue2}).
A simple property we want to prove is that given a list segment representing a
non-empty queue at the beginning, after each operation, we still get back a list segment.
 
In the two use cases, the ``moving pointers'' 
are necessary to recurse differently:
the {\tt tail} is moved in {\tt enqueue} while
the {\tt head} is moved in {\tt dequeue}.
Consequently, no matter how we define list segments\footnote{Typically, 
list segment can be defined in two ways:
the moving pointer is either the \emph{left} one or the \emph{right} one.}, 
where {\tt head} and {\tt tail} are the two pointers, 
at least one use case would recurse differently from 
the definition, thus exhibit the ``recursion divergence'' 
scenario and lead to a proof obligation involving
unmatchable predicates.
More concretely, if list segment is defined as in Fig.~\ref{fig:intro:def},
the {\tt enqueue} operation would lead to an obligation that is
impossible for U+M to prove.

Next, we move along to ``generalization of predicate''.
This happens in almost all realistic programs.
The reason is because verification of functional 
correctness is performed \emph{modularly}.
More specifically, given the specifications for functions and invariants for
loops, we can first perform \emph{local} reasoning before
composing the whole proof for the program using,
in the context of Separation Logic, the 
\emph{frame rule}~\cite{reynolds03course}.
It can be seen that, given such divide-and-conquer strategy,
at the \emph{boundaries} between local code fragments, 
we would need ``generalization of predicate''.
A particularly important relationship between predicates,
at the boundary point, is simply that one (the consequent) 
is \emph{more general} than the other (the antecedent),
representing a valid abstraction step.

\vspace{2mm}
\begin{figure}[htb]
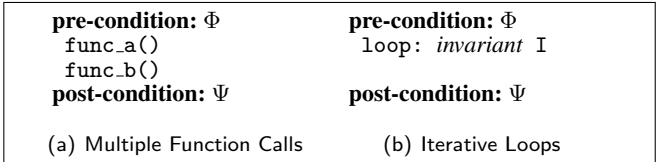

\centering
\subfigure[Multiple Function Calls] {
\mystuff
{
{\bf pre-condition: $\Phi$ ~~~~~~~~~~~~} \\
{ func\_a() } \\
{ func\_b() } \\
{\bf post-condition: $\Psi$} \\
}
\label{fig:intro:function}
}
\hspace{2mm}
\subfigure[Iterative Loops] {
\mystuff
{
{\bf pre-condition: $\Phi$ ~~~~~~~~~~~~} \\
{  loop:} {\it invariant} {I} \\
\\
{\bf post-condition: $\Psi$} \\
}
\label{fig:intro:loop}
}
\vspace{-2mm}
\caption{Modular Program Reasoning}
\label{fig:intro:composition}
\end{figure}

\noindent
Consider the boundaries between function calls, 
illustrated by the pattern in Fig.~\ref{fig:intro:function}.
We start with the pre-condition $\Phi$, calling function {\bf {\tt func\_a}} and then {\bf {\tt func\_b}}.
We then need to establish the post-condition $\Psi$. In traditional forward reasoning, 
we will write local (and consistent) specifications for {\bf {\tt func\_a}} and {\bf {\tt func\_b}} such that:
(1) $\Phi$ is stronger than the pre-condition of  {\bf {\tt func\_a}}; 
(2) the post-condition of  {\bf {\tt func\_a}} is stronger than the pre-condition of  {\bf {\tt func\_b}};
(3) the post-condition of  {\bf {\tt func\_b}} is stronger than $\Psi$. 
It is hard, if not impossible, to ensure that for
each pair (out of three) identified above, the antecedent and the consequent are 
constructed from matchable predicates.
As a concrete example, in {\tt bubblesort} program~\cite{nguyen10shape3},
a boundary between two function calls requires us to prove 
that a \emph{sorted} linked-list is also a linked-list.

We further argue that in software development, code reuse is often desired. 
The specification of a function, especially when it is a \emph{library} function,
should (or must) be relatively \emph{independent} of the context
where the function is plugged in. In each context, we might want to establish arbitrarily different
properties, as long as they are \emph{weaker} than what the function can guarantee.
In such cases, it is almost certain that we will have proof obligations involving
unmatchable predicates.

Now consider the boundaries caused by loops.
In \emph{iterative} algorithms, the loop invariants must be consistent with the code,
and yet these invariants are only used later to prove a property 
often \emph{not} specified using the identical predicates of the invariants.
In the pattern shown by Fig.~\ref{fig:intro:loop},
this means that the proof obligations relating the pre-condition $\Phi$ to the invariant {\tt I}
and {\tt I} to post-condition $\Psi$ often involve unmatchable predicates.  
For example, programs manipulate lists usually have loops of which
the invariants need to talk about list segments. Assume
that (acyclic) linked-list is defined as below:

\vspace{2mm}
\begin{figure}[h!]
\begin{center}
\mystuff{
\listn(x) $\eqdef$  x$=$null $\wedge$ {\bf emp}\\
\> \>  ~       $|$        $\one{\tt x}{\tt t}$ \hsep ~\listn(t) \\
}
\end{center}
\vspace{-6mm}
\end{figure}

\noindent
Though \lsegleft{} and \listn{}
are closely related, U+M can prove neither
of the following obligations:
\numberwithin{equation}{section}
\begin{align}
\lsegleftmath{\tt (x, null)} \models \listn{\tt (x)} \label{ob:1}\\
\lsegleftmath{\tt(x,y)} * \listn{\tt(y)} \models \listn{\tt(x)} \label{ob:2}
\end{align}

\noindent
%
%
%
%
In summary, the above discussion connects to a serious issue in 
software development and verification: 
without the power to relate predicates --- when they are unmatchable --- 
\emph{compositional reasoning} is seriously hampered.
As a matter of fact, the state-of-the-art is only effective for automatically proving a
small fraction of academic and real-world programs.

\subsection*{On using Axioms and Lemmas}

We have argued the necessity of proving relationship between
unmatchable predicates.
Therefore, there cannot be a significant class of programs
that is automatically provable using U+M only.  How is that 
there are in fact existing systems displaying proofs of
significant examples?

One reason is that some existing methods, e.g.,~\cite{cook11concur,
navarro11pldi,navarro13aplas,wies13cav},
only allow properties to be constructed from 
a \emph{pre-defined} set of recursive predicates so that
hard-wired rules can then be used to facilitate U+M. 
For systems that support general user-defined 
predicates~\cite{nguyen10shape3,qiu13dryad}, 
they get around the limitation of U+M via the use, without proof,
of additional \emph{user-provided} ``lemmas'' 
(the corresponding term used in~\cite{qiu13dryad} is ``axioms'').
The general idea is two-fold:
\begin{itemize} 
\item these lemmas have proofs, though manual, which are simple;
\item these lemmas are general and the number of needed ones is small.
\end{itemize}

\noindent
Many lemmas used by \cite{qiu13dryad,pek14pldi}, the most comprehensive existing systems, 
do not satisfy the two conditions.  
Some are not so obvious (therefore not reasonable to accept as proven) 
and some are specifically tailored for the
proofs of the target programs (e.g., the lemma used to prove {\tt delete\_iter} method 
in a binary search tree).

As a matter of fact, it is unacceptable that in order to prove more programs,
we \emph{continually} add in more custom lemmas to facilitate the proof system.
Motivated by this, i.e., to avoid using unproven lemmas,
``Cyclic Proof''~\cite{cyclic11cade,cyclic12aplas} has recently emerged as an advanced proof technique in
Separation Logic. This approach goes beyond U+M and is able 
to prove the relationship between
some commonly used \emph{shape} predicates, even though
they are unmatchable. 

In certain benchmarks, we concern only with properties involving the 
``shape'' of the data structures, e.g., the  obligations \eqref{ob:1} and \eqref{ob:2} above.
In this regard, ``Cyclic Proof'' would be quite effective.
In practice, however, what is often needed are
\emph{custom} predicates for specific application domains, and these
will involve \emph{pure} constraints to capture properties of the 
\emph{data values} in addition to the shapes.
Since generally ``Cyclic Proof'' does not work in the presence of pure constraints,
such obligations are out of its scope.

In short, as a proof method, ``Cyclic Proof'' has advanced the U+M paradigm,
but  it is still too restrictive to be applicable for the purpose of program verification.
We elaborate more in Section~\ref{sec:example}.


\subsection*{Our Contributions}

In this paper, 
we propose a general proof method that goes significantly 
beyond the state-of-the-art,  namely U+M and ``Cyclic Proof'',
therefore being able to prove
relationships between general predicates 
of arbitrary arity, even when recursive definitions and 
the code are structurally dissimilar.
Our  proof method, in one hand, is built upon U+M and therefore subsumes the 
power of existing (un)-fold reasoners.
%
%

On the other hand, in the process of U+M --- \emph{systematically} unfolding
with the hope to reduce the original obligation
to some obligation(s) provable by simple matching and constraint solving ---
we look for a closely similar obligation in the history of one proof path, of which
the truth would \emph{subsume} or \emph{simplify} the current obligation. 
When such ancestor, called an induction hypothesis, is found, 
the current obligation can be re-written into a simpler obligation, 
hopefully then U+M can be effective.
Similar to ``Cyclic Proof", the soundness of such re-writing steps 
are based on the theory of induction~\cite{jaffar08coind}.
In Section~\ref{sec:example}, we will demonstrate in details the limitation of ``Cyclic Proof'' 
as well as why we can consider it as a special and simple realization of the proof rules
presented in this paper.

\ignore{
Importantly, more than one re-writing steps might be needed
before a proof path can be terminated (see Example~\ref{}).
}


Specifically,  this paper makes the following contributions:

\begin{itemize} 
\item We close the remaining reasoning \emph{gaps} in existing state-of-the-art
systems. In particular, we demonstrate in Section~\ref{sec:experiment} that
our proof rules can automatically and efficiently discharge 
all commonly-used lemmas, where existing proof techniques
fail to do so.

\item We also demonstrate on a comprehensive set of benchmarks,
collected from existing systems, that with our proof method the usage of lemmas 
can be \emph{eliminated}. The impact of this is twofold. First,
it means that for proving practical (but small) programs,
the users are now free from the burden of providing custom user-defined lemmas.
Second, it significantly boosts up the performance, since lemma applications,
coupled with unfolding, often induce very large search space.

\item The proposed proof method gets us back the power of compositional reasoning
in dealing with user-defined recursive predicates. While we 
have not been able to identify \emph{precisely} the class where our proof method would 
be effective\footnote{This is as \emph{hard} as identifying the class where an invariant
discovery technique guarantees to work.}; we do believe that
its potential impact is huge.
One important subclass that we can handle effectively is when 
both the antecedent and the consequent refer to the same structural shape 
but the antecedent simply makes a \emph{stronger} statement 
about the values in the structure (e.g., to prove
that a sorted list is also a list, an AVL tree is
also a binary search tree, a list consists of all data values $999$ 
is one that has all positive data, etc.).
 
\end{itemize}

\ignore{
Importantly, in the domain of data structures, the recursive
definitions \emph{are often} structurally similar.  For this reason,
we are able to perform automatic proofs on real and commonly used
benchmarks, that previously required manual assistance.  Though the
algorithm is incomplete (this is necessarily so because the recursion
rules are general), our algorithm often succeeds because there is
often a relationship between the antecedent and consequent of the
proof obligation.  One important example of this is when they both
refer to the same data structure but antecedent simply makes a
stronger statement about the values in the structure (e.g. to prove
that a sorted list is also a list).  As a result of exploiting
structural similarity, we now have a specific feature that we can
automatically prove relationships between different predicates, and of
arbitrary arity.  In summary, we provide a systematic proof method
that significantly extends the state-of-the-art in automatically
proving properties of recursive data structures.
}



\ignore{
One direct implication of this is that we have closed the gaps left by many existing works. 
The potential of automatic induction, however, is not limited to this purpose. 
Indeed, the concept enables unbounded number of automatic and justified
abstraction steps in the search process; therefore we believe our work
not only provides a new level of automation for program
verification but also can act as a basis for an \emph{analysis}
framework, which we will consider as our future work.
}



\section{The State-of-the-art and Motivating Examples}
\label{sec:example}

\noindent
In this Section, we use illustrative examples
to position our proof method against the state-of-the-art.

\subsection*{Unfold-and-Match (U+M) Paradigm}

As stated in Section~\ref{sec:intro},
the dominating technique to manipulate user-defined recursive predicates
is to employ the basic transformation steps of folding and unfolding the rules,
together with formula abstraction, i.e., the U+M paradigm.

The main challenge of the U+M paradigm is clearly how 
to systematically search for such sequences 
of fold/unfold transformations. We believe recent 
works~\cite{madhusudan12dryad,qiu13dryad}, 
we shall call the \dryad{} works, have brought the U+M 
to a new level of automation. 
The key technical step is to use the \emph{program statements} in order to \emph{guide} 
the sequence of fold/unfold steps
of the recursive rules which define the predicates of interest.
For example, assume the definition for list segment~\lsegleft ~in Fig.~\ref{fig:intro:def} 
and the code fragment in Fig.~\ref{fig:ex:lseg1}.

\begin{figure}[h]
\centering
\subfigure[Code Fragment 1] {
\mystuff
{
{ \lsegleft(x,y)}  \\
\small{ ~~assume(x != null) } \\
\small{ ~~z = x.next } \\
{ \lsegleft(z,y)} 
}
\label{fig:ex:lseg1}
}
\hspace{-1mm}
\subfigure[Code Fragment 2] {
\mystuff
{
{ \lsegleft(x,y) $\hsep ~\one{{\tt y}}{\_}$} \\
 \\
\small{ ~~z = y.next } \\
{ \lsegleft(x,z)} 
}
\label{fig:ex:lseg2}
}
\vspace{-2mm}
\caption{U+M with List Segments}

\end{figure}

\begin{figure*}[htb]
\begin{center}
${
\inferrule* [Right=({\tt \lsegleft$(x,y)$})]
  {
  {\inferrule* [Right=]
    {\inferrule* [Right=]
      {\tt True }
      {x{=}y ~{\wedge}~ {\bf emp} ~{\hsep}~ \listn(y) ~{\models}~ \listn(x)}
    }
    {}
  }
  \\  
  {\inferrule* [Right=] 
    {\inferrule* [Right=(*)]
      {{\inferrule* [Right=]
        {\tt True }
        {x{\neq}y ~{\wedge}~ \one{x}{t} ~{\models}~ \one{x}{t}}
      }
      \\ {(\dagger)~ \lsegleftmath(t,y) ~{\hsep}~ \listn(y) ~{\models}~ \listn(t)}}
      {x{\neq}y ~{\wedge}~ \one{x}{t} ~{\hsep}~ \lsegleftmath(t,y) ~{\hsep}~ \listn(y) ~{\models}~ \one{x}{t} ~{\hsep}~ \listn(t)}
    }
    {x{\neq}y ~{\wedge}~ \one{x}{t} ~{\hsep}~ \lsegleftmath(t,y) ~{\hsep}~ \listn(y) ~{\models}~ \listn(x)}
  {}
  }
  }
  {(\dagger)~ \lsegleftmath(x,y) ~\hsep ~\listn(y) ~\models~ \listn(x)}
}$
\end{center}
\caption{Proving with Cyclic Proof}
\label{fig:ex:cyclic}
\end{figure*}

\noindent Here we want to prove that given \lsegleft{\tt(x,y)} at the beginning, 
we should have \lsegleft{\tt(z,y)} at the end.
Since the code touches the ``footprint'' of {\tt x} (second statement), it \emph{directs} the unfolding of
the predicate containing {\tt x}, namely \lsegleft{\tt (x,y)}, to expose
${\tt  x \neq y \wedge \one{x}{t}~\hsep}$~\lsegleft{\tt(t,y)}. The consequent  
can then be established via a simple \emph{matching} from variable {\tt z} to {\tt t}.

Now we consider the code fragment
in Fig.~\ref{fig:ex:lseg2}: instead of moving one position away
from {\tt x}, we move one away from {\tt y}. To be convinced 
that U+M, however, cannot work,
it suffices to see that unfolding/folding of \lsegleft{}
does not change the \emph{second argument} of the predicate \lsegleft.
Therefore, regardless of the unfolding/folding sequence, the arguments {\tt y} on the LHS 
and {\tt z} on the RHS
would maintain and can never be matched satisfactorily.

The example in Fig.~\ref{fig:ex:lseg2} exhibits the ``recursion divergence'' 
scenario mentioned in Section~\ref{sec:intro} and ultimately is about relating 
two possible definitions of list segment (recursing either on the left or on the right pointer), 
which U+M fundamentally cannot handle.
We will revisit this example in later Sections.

\subsection*{Cyclic Proof}

``Cyclic Proof''~\cite{cyclic11cade,cyclic12aplas} method operates via the key
observation that when two similar obligations are detected in the same 
proof path, the latter can be used to subsume the former. 
The soundness is based on the theory of induction.
In other words, ``Cyclic Proof'' can soundly terminate a proof path if the
current obligation can be derived from some ancestor obligation 
in the history of the proof path, just by variable renaming. 
Of course, we do need to ensure a progressive measure in order
to implement ``Cyclic Proof''. However, for simplicity,
we will omit such details in this discussion.

We can view ``Cyclic Proof'' as a significant extension of U+M.
Other than terminating when all the predicates in the RHS can be matched
satisfactorily with some predicates in the LHS, as in U+M, 
it allows termination of a proof path when the \emph{whole} current obligation 
can be matched with some (whole) obligation appeared in the history
of the path. Let us illustrate with a concrete example.

\begin{example}
Consider the obligation \lsegleft{\tt(x,y)} * \listn{\tt(y)} $\models$ \listn{\tt(x)},
i.e., obligation \eqref{ob:2} introduced in Section~\ref{sec:intro}.
\end{example}

\noindent
See Fig.~\ref{fig:ex:cyclic}, where the \emph{cyclic path} is at rightmost position.
We will focus on this interesting path.
The proof proceeds by first unfolding the definition of~\lsegleft{\tt (x,y)} in the LHS, 
followed by unfolding the definition of ~\listn{\tt (x)} in the RHS.
The separation rule ($\hsep$) allows us to simplify and subsequently observe
the cyclic path, with the appropriate pair denoted with $(\dagger)$.

``Cyclic Proof'' does open avenue for the integration of user-defined 
recursive definitions and auxiliary lemmas that relate such definitions, 
but ``it apparently cannot handle definitions where \emph{pure} constraints are present''~\cite{stewart12icfp}.
In other words, it can only help reasoning about the shapes of data structures; but
it is not effective when  properties of the data values are involved.

Specifically, ``Cyclic Proof'' is not able to prove the extended versions of 
obligations \eqref{ob:1} and \eqref{ob:2}, which are used frequently in \cite{qiu13dryad},
when we also want to establish the relationship of 
collective data values (using sets or sequences) between the LHS and the RHS.
It can neither prove that a \emph{sorted} linked-list is also
a linked-list (in verifying {\tt bubblesort} program) 
nor reason about the relationships between \emph{sorted} list segments and \emph{sorted} lists,
which are necessary to verify {\tt quicksort} and {\tt mergesort} programs~\cite{qiu13dryad}.

Also importantly, ``Cyclic Proof'' cannot handle any obligation
of which the LHS and the RHS exhibit \emph{recursion divergence}.
For example, it will not be able to prove either
\lseg{\tt(x,y)} * \listn{\tt(y)} $\models$ \listn{\tt(x)} or
\lsegleft{\tt(x,y)} $\models$ \lseg{\tt(x,y)}\footnote{Such 
obligations arise often in real programs
dealing with lists (e.g., the library {\bf glib/gslist.c}).}, 
where \lseg, another form of list segment,  is defined as follows.

\vspace{1mm}
\begin{figure}[h!]
\begin{center}
\mystuff{
\lseg(x,y) $\eqdef$  x$=$y $\wedge$ {\bf emp}\\
\> \>  \>       $|$        x$\neq$y $\wedge$ $\one{\tt t}{\tt y}$ \hsep ~\lseg(x,t) \\
}
\end{center}
\vspace{-6mm}
\end{figure}
\vspace{1mm}

\noindent
The reason is that, with recursion divergence, unfolding will typically 
introduce \emph{new} existential variable(s) on the RHS, which cannot be 
matched with any variables in the LHS.
As a result, (the RHS cannot be simplified and) we will not observe a 
pair of ancestor-descendant obligations, 
whose difference is only by variable renaming.

The key distinction that enables us to handle such obligations is that
our method treats previous obligations in a proof path as possible (dynamic) induction 
hypotheses and therefore allows the current obligation not only to be terminated, but also 
to be \emph{re-written} into a new obligation, usually simpler. The soundness
of each rewriting step is also based on the theory of induction.
In Section~\ref{rules:motivating}, we demonstrate our method on these two obligations.

Since  ``Cyclic Proof'' will terminate a proof path 
and declare the current obligation as proven
when it finds a ``look-alike'' ancestor obligation --- 
achieved by renaming variables on the current obligation,
therefore, it can be considered a \emph{special} and simple
realization of our proof rules.
One can view ``Cyclic Proof'' to our proof method analogously as 
cycle detection in explicit model checking to the process 
of widening\footnote{Technically,
in the context of proving, our re-writing step is a narrowing operation.} 
in order to discover an invariant with the hope to terminate the reasoning
of an unbounded loop.

\floatstyle{plain}
\restylefloat{figure}

\vspace{3mm}
\begin{figure}[h]
\begin{center}
\begin{tikzpicture}
\draw (0.2, 0) ellipse [x radius=0.7cm, y radius=1.5cm, rotate=90] node [label={[xshift=-0.4cm, yshift=-0.3cm]U+M}] {};
\draw (1.6, 0) ellipse [x radius=0.3cm, y radius=0.8cm, rotate=90] node [label={[xshift=0.4cm, yshift=-0.3cm]CP}] {};
\draw (1.6, 0) ellipse [x radius=1.0cm, y radius=3.5cm, rotate=90] node [label={[xshift=2.2cm, yshift=-0.3cm]Our Proof Method}] {};
\end{tikzpicture}
\end{center}
\caption{Correlation between U+M, CP, and our proof method.}
\label{venn}
\end{figure}
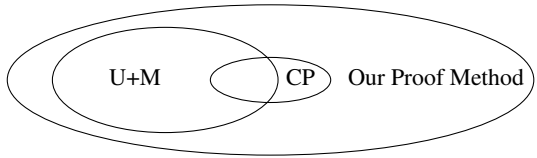

\floatstyle{boxed}
\restylefloat{figure}

\vspace{2mm}
\noindent To summarize, in Fig.~\ref{venn}, we illustrate the correlation
between U+M paradigm, ``Cyclic Proof" (CP) method, and our proof method.
Theoretically, ``Cyclic Proof" subsumes U+M techniques, due to the capability
of reasoning about unmatchable predicates. However, in its current implementation,
``Cyclic Proof" does not support general non-shape properties
(as shown in the {\sf SMT-COMP 2014} for Separation Logic). 
Since such properties (e.g., size property)
are ubiquitous in practical programs, there is a large portion of proof obligations 
that U+M can prove but ``Cyclic Proof" cannot. 
For the domain of interest, our proof method, 
in contrast, subsumes 
both the U+M paradigm and ``Cyclic Proof" technique. 
We later demonstrate this clearly in our Experimental Evaluation.

%
%
%
%
%
%

\subsection*{Automatic (and Explicit) Induction}

In the literature, there have been works on automatic induction~\cite{ACL1,ACL2,leino12vmcai,ZENO}. 
They are concerned with proving a \emph{fixed} hypothesis, say $h(\tilde{x})$, that is, to show
that $h()$ holds over all values of the variables $\tilde{x}$.
The challenge is to discover and prove
$h(\tilde{x})~{\implies}~h(\tilde{x}')$, where expression $\tilde{x}$ 
is less than the expression $\tilde{x}'$  in some well-founded measure.
Furthermore, a \emph{base case} of $h(\tilde{x}_0)$ needs to be proven.
Automating this form of induction usually relies on the fact that
some subset of $\tilde{x}$ are variables of \emph{inductive types}.

In contrast, our notion of induction hypothesis is completely different.
First, we do not require that some variables are of inductive (and well-founded) types.
Second, the induction hypotheses are not supplied explicitly.
Instead, they are constructed implicitly via the discovery of a valid proof path.
This allows much more potential for automating the proof search.
Third, and also different from ``Cyclic Proof'', multiple induction hypotheses
can be exploited within a single proof path. Without this, we would not be able to prove
\lsegleft{\tt(x,y)} $\models$ \lseg{\tt(x,y)}.


\ignore{
We now present a few examples which we can prove automatically,
and for which the current state of the art \cite{nguyen10shape3,qiu13dryad}
cannot.  These examples demonstrate very basic program fragments
and properties.  Therefore the kind of reasoning used here is likely to be
quite widely applicable.

The first example concerns a queue, implemented as a doubly-linked list.
The predicate $queue(H, head, tail, F)$ below simply states that there
is a doubly-linked list from $head$ to $tail$, and its footprint is $F$.
Note that we assume the offsets of the pointer fields $next$ and $prev$
are $+1$ and $+2$ respectively.
The following programs takes a queue and dequeues and enqueues it.
The interesting aspect here is that the data structure has two handles,
and that the programs need to move both of them.
Thus whichever way we choose to specify the $queue$ predicate,
whether by tail-recursion from the LHS (as in the example), or by recursing
from the RHS, the recursion direction will be different from
one of the two program fragments.

\medskip
\noindent\fbox{
\begin{minipage}{\columnwidth}
{\fontsize{8}{10}\selectfont
\stuff{
queue($H, head, tail, F$) :- $head = 0, tail = 0, F \heq \hemp$. \\
queue($H, head, tail, F$) :- \\
\> $head \neq 0, head = tail$, \\
\> $H[head + 1] = 0, H[tail + 2] = 0, F \heq (head \mapsto \_)$. \\
queue($H, head, tail,  F$) :- \\
\> $head \neq 0, head \neq tail,$ \\
\> $H[head + 1] = tmp, H[tmp + 2] = head$, \\
\> $F \heq (head \mapsto \_) * F1$, queue($tmp, tail, F1$). \\

}
}
\end{minipage}
}
\medskip

\noindent\fbox{
\begin{minipage}{\columnwidth}
{\fontsize{8}{10}\selectfont
\stuff{
$Requires: queue(H, head, tail, \_)$ \\
$Ensures: queue(H, head, tail, \_)$ \\
\>        if (head == null) return; \\
\>            if (head == tail) \{ \\
\>\>               x = head; \\
\>\>               head = null; tail = null; \\
\>            \} else \{ \\
\>\>              x = head; \\
\>\>              head = head${\rightarrow}$next; head${\rightarrow}$prev = null; \\
\>            \}\\
}
}
\end{minipage}
}

\medskip
\noindent\fbox{
\begin{minipage}{\columnwidth}
{\fontsize{8}{10}\selectfont
\stuff{

$Requires: queue(H, head, tail, F), (x \mapsto \_) * F$ \\
$Ensures: queue(H, head, tail, \_)$ \\
\>        if (tail == null) \{\\
\>\>          head = tail = x; head${\rightarrow}$next = head${\rightarrow}$prev = null; \\
\>       \} else \{\\
\>\>            x${\rightarrow}$next = null; x${\rightarrow}$prev = tail; \\
\>\>            tail${\rightarrow}$next = x; tail = x;\\
\>        \}\\
}
}
\end{minipage}
}

\vspace*{3mm}
\noindent
In the next example, the predicate $successor(H, x, y)$ specifies two \emph{different}
ways that $x$ and $y$ can be related to each other.
In what follows, think of $y = H[x +1]$ as indicating
that $y$ is a mother of $z$, while $y = H[x +2]$ means $y$ is the father.
Proving the (one-line) program is problematic for DRYAD
because its ``unfolding over the footprint'' of the program
does not correspond to a unique unfolding of the predicate.
We also note here that the work \cite{iosif13treewidth} which represents a recent
and relatively general decidable logic for heaps and separation,
cannot accommodate this example.

\medskip
\noindent\fbox{
\begin{minipage}{\columnwidth}
{\fontsize{8}{10}\selectfont
\stuff{
successor($H, x, y$) :- $y = H[x + 1]$.\\
successor($H, x, y$) :- $y = H[x + 2]$.\\
samelevel($H, x, y$) :- $x = y$.\\
samelevel($H, x, y$) :- \\
\> sucessor(H,z,x), succesor(H,t,y), samelevel(H,z,t).\\
\ \\
$Requires: successor(H, x, y), x \neq 0$ \\
$Ensures: samelevel(H, x, y)$ \\
\> x = x${\rightarrow}$left; 
}
}
\end{minipage}
}

\vspace*{3mm}

\noindent
The next example extends deals with a given array segment of zero elements, and extends 
the segment by one.  The interesting point here is that the recursion in the 
predicate is ``tail-recursive'', while the program corresponds to a
definition which is ``segment'' recursive, that is, 
DRYAD would be able to prove this example if the recursive rule below
used $zeroseg($H, m, n-1$)$ instead of $zeroseg($H, m+1, n$)$.

\medskip
\noindent\fbox{
\begin{minipage}{\columnwidth}
{\fontsize{8}{10}\selectfont
\stuff{
zeroseg($\_, m, n$) :- $m > n$. \\
zeroseg($H, m, n$) :- $m \leq n, H[m] = 0$, zeroseg($H, m+1, n$). \\
\ \\
$Requires: zeroseg(H, m, n)$ \\
$Ensures: zeroseg(H, m, n)$ \\
\> [m+1] = 0; \\
\> m = m + 1;
}
}
\end{minipage}
}

\vspace*{3mm}
\noindent
Finally, we present a non-heap example.
In this example, the predicate $fib(x, y)$ specifies that $y$ is the $x^{th}$ Fibonacci number
in the traditional recursive manner, while the program implements the familiar
iterative algorithm.  The arising VC of interest here is that the $Invariant$ below is indeed
a loop invariant.

\vspace*{3mm}
\noindent\fbox{
\begin{minipage}{\columnwidth}
{\fontsize{8}{10}\selectfont
\stuff{
fib($n, 1$) :- $0 \leq n \leq 1$. \\
fib($n, m1+m2$) :- $n > 1$, fib($n-1, m1$) + fib($n-2, m2$). \\
\ \\
$Invariant: fib(i-1, a), fib(i, b)$ \\
\> a = b = i = 1; \\
\> while (i < n) \{ \\
\>\> tmp = b; b = a + b; a = tmp; i = i+1 \\
\> \}
}
}
\end{minipage}
}

\ignore{
{\fontsize{8}{10}\selectfont
\stuff{
lseg-even($x,y,{x}*{y}$) :- $x \neq y, y \neq nil, x{\rightarrow}next = y$. \\
lseg-even($x,y,\{z\}*\{t\}*F$) :- $y \neq nil, x{\rightarrow}next = z$, \\
\>\>              $t{\rightarrow}next = y, lseg-even(z,t,F)$ \\ 
\ \\
$Requires: lseg-even(x, y, F)$ \\
$Ensures:  lseg-even(x, y, F)$ \\
\>    z = malloc()\\
\>    t = malloc()\\
\>    z${\rightarrow}$next = x;\\
\>    y${\rightarrow}$next = t;\\
\>    x = z;\\
\>    y = t;\\

}
}

\note{CADE13 cannot handle this}

{\fontsize{8}{10}\selectfont
\stuff{
    lseg-even($x,y,{x}*{y}$) :- $x \neq y, y \neq nil, x{\rightarrow}next = y$. \\
    lseg-even($x,y,\{z\}*\{t\}*F$) :- $y \neq nil, x{\rightarrow}next = z$, \\
\>\>              $t{\rightarrow}next = y, lseg-even(z,t,F)$ \\ 
}
}

{\fontsize{8}{10}\selectfont
\stuff{
    requires: lseg(x,y,F), z $\not \in$ F, t $\not \in$ F \\
    ensures: lseg(z,t,\{z\}*\{y\}*F) \\
    assume(z != nil);\\
    asumme(y != nil);\\
    z.next = x;\\
    y.next = t;\\
}
}

{\fontsize{8}{10}\selectfont
\stuff{
    lseg1(x,y,emp) :- x = y. \\
    lseg1(x,y,\{t\}*F) :- x != y, lseg1(x,t,F), t.next = y.\\

    lseg2(x,y,emp) :- x = y.\\
    lseg2(x,z,\{x\}*F) :- x != y, z = x.next, lseg2(z,y,F).\\

    lseg1(x,y,F) |= lseg2(x,y,F).\\
    lseg2(x,y,F) |= lseg1(x,y,F).\\
}
}
}
}

\section{The Assertion Language}
\label{sec:prelim}
The explicit naming of heaps has emerged naturally in
several extensions of Separation Logic (SL) as an aid to practical program
verification. Reynolds conjectured that referring explicitly to the
current heap in specifications would allow better handles on data structures 
with sharing \cite{reynolds03course}.
Duck et al. \cite{duck13heaps}, in this vein, extends Hoare Logic with explicit heaps.
This extension allows for strongest post conditions, and is therefore 
suitable for ``practical program verification'' \cite{jules14popl} via constraint-based symbolic execution.

In this paper, we start with the existing specification language
in  \cite{duck13heaps}, which has two notable features: (a) the use of
explicit heap variables, and (b) user-defined recursive properties in
a wrapper logic language based on recursive rules.  
The language provides a new level of
expressiveness for specifying properties of heap-manipulating programs.
We remark that, common specifications written in traditional Separation Logic, 
can be automatically compiled into this language.

Due to space limit, we will be brief here and refer interested readers to~\cite{duck13heaps}
for more details.
A \emph{heap} is a \emph{finite partial map} from \emph{positive}  integers to integers, i.e.
$\set{Heaps} = \Z_{+}  \parmap_{\text{fin}}\Z$.
Given a heap $h \in \set{Heaps}$ with domain $D = \dom(h)$,
we sometimes treat $h$ as the set of pairs
$\{(p, v)~|~p \in D \wedge v = h(p)\}$.
We note that when a pair $(p, v)$ belongs to some heap $h$,
it is necessary that $p$ is not a {\tt null} pointer, i.e., $p \neq 0$.
The $\HH$-language is the first-order language over heaps.

We use ($\hsep$) and ($\heq$) operators to respectively denote heap disjointness 
and equation. Intuitively, a constraint like $H \heq H_1 \hsep H_2$ 
restricts $H_1$ and $H_2$ to be disjoint while 
giving a name $H$ to the conjoined heaps
$H_1 \hsep H_2$.

\subsection*{The Assertion Language $CLP(\HH)$}

As in~\cite{duck13heaps}, $\HH$ is then extended with \emph{user-defined} recursive predicates.
We use the framework of \emph{Constraint Logic Programming} 
(CLP)~\cite{jaffar94clp} to inherit its syntax, semantics, and most importantly, its
built-in notions of unfolding rules.
For the sake of brevity, we just informally explain the language.
The following rules constitutes a recursive definition 
of predicate $\listn(x, L)$, which
specifies a skeleton \emph{list}.

\vspace*{3mm}
\stuff{
\listn($x, L$) \clpif~ $x = 0$, $L \heq \hemp$. \\
\listn($x, L$) \clpif~ $L \heq \one{x}{t} \hsep L_1$, \listn($t, L_1$).
}
\vspace*{3mm}

\noindent
The \emph{semantics} of a set of rules is traditionally 
known as the ``least model'' semantics (LMS).
Essentially, this is the set of groundings of the predicates which are true
when the rules are read as traditional implications.
The rules above dictates that all true groundings of $\listn(x, L)$ are such that $x$ is
an integer, $L$ is a heap which contains a skeleton list starting from $x$.
More specifically, when the list is empty, the root node is equal to {\tt null} ($x = 0$),
and the heap is empty ($L \heq \hemp$).
Otherwise, we can split the heap $L$ into two disjoint parts:
a singleton heap $\one{x}{t}$ and the remaining heap $L_1$,
where $L_1$ corresponds to the heap that contains a skeleton list starting from $t$.

We now provide the definitions for list segments, which will
be used in our later examples. 
Do note the extra explicit heap variable $L$, in comparison
with corresponding definitions in SL.

\vspace*{3mm}
\stuff{
\lsegleft($x, y, L$) \clpif~ $x = y$, $L \heq \hemp$. \\
\lsegleft($x, y, L$) \clpif~ $x \neq y$, $L \heq \one{x}{t} \hsep L_1$, \lsegleft($t, y, L_1$).
}
\vspace*{3mm}

\stuff{
\lseg($x, y, L$) \clpif~ $x = y$, $L \heq \hemp$. \\
\lseg($x, y, L$) \clpif~ $x \neq y$, $L \heq \one{t}{y} \hsep L_1$, \lseg($x, t, L_1$).
}
\vspace*{3mm}

\noindent
We also emphasize that the main advantage of this language is the possibility
of deriving the strongest postcondition along each program path.
It is indeed the main contribution of~\cite{duck13heaps}.
Specifically, in order to prove the Hoare triple 
$\{\phi\} S \{\psi\}$ 
for a loop-free program $S$, we simply generate strongest postcondition $\psi'$
along each of its straight-line paths and obtain the verification condition
$\psi' \models \psi$.
Note that the handling of loops can be reduced to this loop-free
setting because of user-specified invariants.
We put forward that, in all our experiments (Section~\ref{sec:experiment}), the verification conditions 
are generated using the symbolic execution rules of~\cite{duck13heaps}.




\section{The Proof Method}
\label{sec:proofrules}
\label{proofmethod}

{\bf Background on CLP:} This is provided for the convenience of the readers.
An {\em atom} is of the form $p(\tilde{t})$ where $p$ is a
user-defined predicate symbol and $\tilde{t}$ is a tuple of $\HH$ terms.
A {\em rule\/} is of the form $A \clpif \Psi, \tilde{B}$ where the atom
$A$ is the {\em head} of the rule, and the sequence of atoms
$\tilde{B}$ and the constraint $\Psi$ constitute the {\em body} of the
rule.  
A finite set of rules is then used to define a predicate.
A {\em goal} has exactly the same format as the body of a rule. 
A goal that contains only constraints and no atoms is called {\em final}. 

A {\it substitution} $\theta$ simultaneously replaces each variable in
a term or constraint $e$ into some expression, and we write $e\theta$
to denote the result.  A {\it renaming} is a substitution which maps
each variable in the expression into a distinct variable.  
A {\it grounding} is a substitution which maps each 
variable into its intended universe of discourse: an integer or a heap, in the case of our CLP($\HH$).  
Where $\Psi$ is a constraint, a grounding of $\Psi$ results in
\true{} or \false{} in the usual way.

A {\em grounding\/} $\theta$ of an atom $p(\tilde{t})$ is an object of
the form $p(\tilde{t}\theta)$ having no variables. A grounding of a
goal $\G \equiv (p(\tilde{t}), \Psi)$ is a grounding $\theta$ of
$p(\tilde{t})$ where $\Psi\theta$ is \true.  We write $\lbr\G\rbr$ to
denote the set of groundings of $\G$.


Let $\G \equiv (B_1, \cdots , B_n, \Psi)$ and $P$ 
denote a non-final goal and a set of rules respectively.
Let $R \equiv A \clpif ~\Psi_1, C_1, \cdots , C_m$ denote a rule in $P$,
written so that none of its variables appear in $\G$.
Let the equation $A = B$ be shorthand for the pairwise 
equation of the corresponding arguments of $A$ and $B$.
A {\em reduct} of $\G$ using a clause $R$,
denoted $\red(\G,R)$, 
is of the form

        $(B_1, \cdots , B_{i-1}, C_1, \cdots , C_m, B_{i+1}, 
        \cdots, B_n, B_i = A, \Psi, \Psi_1)$

\noindent
provided the constraint $B_i = A \wedge \Psi \wedge \Psi_1$ is satisfiable.

A {\em derivation sequence} for a goal $\G_0$ is a possibly infinite 
sequence of goals $\G_0, \G_1, \cdots$, where $\G_i,~i > 0$ is
a reduct of $\G_{i-1}$. 
A {\em derivation tree} for a goal is defined in the obvious way.

\begin{definition}[Unfold]
  Given a program $P$ and a goal $\G$:\\
  $~~~~~~~~~~\unfold(\G)$ is $\{\G'|\exists R \in P: \G' =\red(\G,R)\}$.  \qed
\end{definition}

\noindent
Given a goal $\Left$ and an atom $p \in \Left$, 
$\unfold_p(\Left)$ denotes the set of formulas transformed from $\Left$
by unfolding $p$.

\begin{definition}[Entailment]
An {\em entailment} is of the form 
$\Left \models \Right$, 
where $\Left$ and $\Right$ are goals.
\qed
\end{definition}

\noindent
This paper considers proving the validity of the entailment 
$\Left \models \Right$ under a given program $P$.
This entailment means that 
$lm(P) \models (\Left \rightarrow \Right)$, where
$lm(P)$ denotes the ``least model'' of the
program $P$ which defines the recursive predicates {---}
called \emph{assertion} predicates {---} occurring in $\Left$ and $\Right$.
This is simply the set of all groundings of atoms of the assertion  
predicates which are \emph{true} in $P$.
The expression $(\Left \rightarrow \Right)$
means that, for each grounding $\theta$ of $\Left$ and $\Right$,
$\Left\theta$ is in $lm(P)$ implies that so is $\Right\theta$.

%

\subsection{Unfold and Match (U+M)}

Assume that we start off with $\Left \models \Right$.
If this entailment can be proved directly, by
unification and/or consulting an off-the-shelf \smt{} solver, 
we say that the entailment is trivial: 
a \emph{direct proof} is obtained even without considering the ``meaning''
of the recursively defined predicates 
(they are treated as {\em uninterpreted}).
When it is not the case {---} the entailment is non-trivial
{---} a standard approach is to 
apply unfolding/folding until all the 
``frontier'' become trivial.
We note that, in our framework, we perform 
only unfolding, but now to both the LHS (the antecedent) and the 
RHS (the consequent) of the entailment. The effect of unfolding the 
RHS is similar to a folding operation on the LHS. 
In more details, when direct proof fails, 
U+M paradigm proceeds in two possible ways:

\begin{itemize}

\item First, select a recursive atom $p \in \Left$,
unfold $\Left$ wrt. $p$ and obtain the goals
$\Left_1, \ldots, \Left_n$. The validity of the original entailment
can now be obtained by ensuring the validity of \emph{all} the
entailments $\Left_i \models \Right ~(1 \leq i \leq n)$.

\item Second, select a recursive atom $q \in \Right$,
unfold $\Right$ wrt. $q$ and obtain the goals 
$\Right_1, \ldots, \Right_m$. The validity of the original entailment
can now be obtained by ensuring the validity of \emph{any one of} the
entailments $\Left \models \Right_j ~(1 \leq j \leq m)$.

\end{itemize}

\noindent
So the proof process can proceed recursively either by proving
\emph{all} $\Left_i \models \Right$ or 
by proving \emph{one} $\Left \models \Right_j$ for some $j$.
Since the original LHS and RHS
usually contain more than one recursive atoms, 
this proof process naturally triggers a search tree.
Termination can be guaranteed 
by simply bounding the maximum
number of left and right unfolds allowed.
In practice, the number of recursive atoms used in an entailment is usually small, 
such tree size is often manageable.

\ignore{
{\sc dryad} works \cite{madhusudan12dryad,qiu13dryad}, considered as (U+M) state-of-the-arts,
contributed by employing the program code in order to 
direct the unfolding process.
In their restricted language for defining assertion predicates, 
the unfolding process usually becomes \emph{deterministic}.
In fact, they did not need to distinguish between the semantics
of left and right unfolding.
}

\subsection{Formula Re-writing with Dynamic Induction Hopotheses}

\begin{figure*}[tb!]
\vspace{-3mm}
\ \\
$$
\begin{array}{rl}
 \mbox{\sc (cp)} & \hspace{-2pt}
\frac
{\raisebox{4pt}{${\tt True}$}} 
{\raisebox{-5pt}{$\tilde{A}\vdash \Left \models \Right$}}
{\footnotesize
        \begin{array}{l}
        ~~\Left \models_{\tt SMT} \Right 
        \mbox{, where recursive atoms are treated as uninterpreted} \\
        \end{array}
}
\\[15pt]
\mbox{\sc (sub)} & \hspace{-2pt}
\frac
{\raisebox{4pt}{$\tilde{A} \vdash \Left \wedge p(\tilde{x}) \models \Right\theta$}}
{\raisebox{-5pt}{$\tilde{A}\vdash \Left\wedge p(\tilde{x})\models \Right\wedge p(\tilde{y})$}}
{\footnotesize
        \begin{array}{l}
\mbox{~there exists a substitution } \theta 
\mbox{ for}\\
\mbox{~existential variables in }
\tilde{y} \mbox{ s.t. } \Left \wedge p(\tilde{x}) \models_{\tt SMT} \tilde{x} = \tilde{y}\theta
        \end{array}
}
\\[15pt]

\mbox{\sc (lu+i)} & \hspace{-2pt}
\frac
{\raisebox{4pt}{$
        \bigcup_{i=1}^{n} \{
                \tilde{A} \cup \{\pair{\Left \models \Right}{p}\} \vdash 
                \Left_i \models \Right 
                \}$}
} 
{\raisebox{-5pt}{$\tilde{A}\vdash \Left \models \Right$}}
        {\footnotesize
        \begin{array}{l}
             \mbox{~~Select an atom } p \in \Left \mbox{~and}\\
             ~~\unfold_{p}(\Left) = 
             \{\Left_1, \ldots , \Left_n\} 
        \end{array}}

\\[15pt]

\mbox{\sc (ru)} & \hspace{-2pt}
\frac
{\raisebox{4pt}{$\tilde{A}\vdash \Left \models \Right'$}}
{\raisebox{-5pt}{$\tilde{A} \vdash \Left \models \Right$}}
{\footnotesize
\begin{array}{l}
        \mbox{ ~~~~Select an atom } q \in \Right \mbox{~and}\\
        ~~~~~\Right' \in \unfold_{q}(\Right)
 \end{array}}
\\[15pt]
 
\mbox{\sc (ia-1)} & \hspace{-2pt}
\frac
{\raisebox{4pt}{$\tilde{A}\vdash \Right'\theta \wedge \Left_2 \models \Right$}}
{\raisebox{-5pt}{$\tilde{A}\vdash p(\tilde{x}) \wedge \Left_1 \wedge \Left_2 \models \Right$}}
{\footnotesize
\begin{array}{l}
~\pair{p(\tilde{y}) {\wedge} \Left'   \models \Right'}{p(\tilde{y})} \in \tilde{A} 
\mbox{ and }{\tt gen}(p(\tilde{x})) {\geq} {\tt kill}(p(\tilde{y})), \\
\mbox{~there exists a renaming } \theta 
\mbox{ s.t. } \tilde{x} = \tilde{y}\theta
\mbox{ and } \Left_1 \models_{\tt DP} \Left' \theta\\
\end{array}} 

\\[20pt]
\mbox{\sc (ia-2)} & \hspace{-2pt}
\frac
{\raisebox{4pt}{$\tilde{A}\vdash 
\Left_1 \models \Left'\theta
$}}
{\raisebox{-5pt}{$\tilde{A}\vdash p(\tilde{x}) \wedge \Left_1 \models \Right$}}
{\footnotesize
\begin{array}{l}
~\pair{p(\tilde{y}) \wedge \Left' \models \Right'}{p(\tilde{y})} \in \tilde{A}
\mbox{ and }{\tt gen}(p(\tilde{x})) \geq {\tt kill}(p(\tilde{y}))\\
\mbox{~and there exists a renaming } \theta \mbox{ s.t. } 
\tilde{x} = \tilde{y}\theta \mbox { and } \Right'\theta \models_{\tt DP} \Right

\end{array}}

\end{array}
$$

\caption{General Proof Rules}\label{fig:proofrules}

\end{figure*}

\ignore{

Reasoning about different predicates is unavoidable in dealing with the iterative nature
of the program code, while the assertion predicates are intuitively defined in a recursive manner.
In such setting, by repetitively performing left and right
unfold will not reduce the proof obligation to trivial 
forms (where direct proofs are applicable).
}

In Section~\ref{sec:intro}, we have identified one major weakness that U+M, 
which is not being able to relate between unmatchable predicates.
Such scenario happens in most realistic programs due to the boundaries
created by the multiple function calls and/or iterative loops. 
We now present a formal calculus for the proof of $\Left \models \Right$
that goes beyond unfold-and-match. The power of our proof framework 
comes from the key concept: {\em induction\/}. 


\begin{definition}[Proof Obligation]
A {\em proof obligation} is of the form 
$\tilde{A} \vdash \Left \models \Right$ 
where $\Left$ and $\Right$ are goals and $\tilde{A}$ 
is a set of pairs $\pair{A}{p}$, where $A$ is an {\em assumed} entailment and
$p$ is a {\em recursive atom}.
\qed
\end{definition}

\noindent
The role of proof obligations is to capture the state of the proof process.
Each element in $\tilde{A}$ is a pair, of which the first is
an entailment $A$ whose truth can be assumed inductively.
$A$ acts as a dynamic induction hypothesis and can be used to transform 
subsequently encountered obligations in the proof path. We will explain this
process later in more details. The second is a recursive
atom $p$, to which the application of a left unfold gives
rise to the addition of the induction hypothesis $A$.

Our proof rules~--~the obligation at the bottom, and its reduced form on top~--~
are presented in Fig.~\ref{fig:proofrules}.
Given $\Left \models \Right$, our proof shall start with
$\emptyset\vdash \Left\models \Right$, and proceed
by repeatedly applying these rules.
Each rule operates on a proof obligation. In this process, the
proof obligation may be discharged (indicated by {\tt True}); or new proof
obligation(s) may be produced.  
$\Left \models_{\tt SMT} \Right$ denotes that the validity of $\Left \models \Right$
can be obtained simply by consulting a generic \smt{} solver.

\noindent\textbullet~~The {\it substitution} {\sc (sub)} rule removes one occurrence 
of an assertion predicate, say atom $p(\tilde{y})$, appearing in the RHS of a proof 
obligation. Applying the {\sc (sub)} rule repeatedly will ultimately 
reduce a proof  obligation to the form 
which contains no recursive atoms in the RHS, while at
the same time (hopefully) most existential variables on the RHS 
are eliminated. Then, the {\it constraint proof} ({\sc cp}) 
rule may be attempted by simply treating
all remaining recursive atoms (in the LHS) as uninterpreted
and by applying the underlying theory solver assumed in the language we use.

The combination of {\sc (sub)} and {\sc (cp)} rules attempts,
what we call, a {\em direct proof}.
In principle, it is similar to the process of ``matching'' in the U+M paradigm.
For brevity we then use $\Left \models_{\tt DP} \Left$ to denote 
the fact that the validity of $\Left \models \Right$ can be proved directly  
using only ({\sc sub}) and ({\sc cp}) rules.

\ignore{
The main difference is that, in their restricted setting, 
finding the appropriate substitution for existential 
variables in the RHS is ``deterministic''.
Here the task requires some amount of search for matching
recursive atoms. However,
the cost of this search is negligible, since the number of 
recursive atoms appear in the proof obligation is small. 
}

\noindent\textbullet~~The {\it left unfold with induction hypothesis} ({\sc lu+i}) is a key rule.
It selects a recursive atom $p$ on the LHS and 
performs a complete unfold of the LHS wrt. the atom $p$,  
producing a new set of proof obligations.  
The original obligation, while being removed,
is added as an assumption to every newly produced proof obligation, 
opening the door for the later being used as an induction hypothesis.
For technical reason needed below, we do not just add the obligation
$\Left \models \Right$ as an assumption, 
but also need to keep track of the atom $p$. 
This is why in the rule we see a pair $\pair{\Left \models \Right}{p}$ added
into the current set of assumptions $\tilde{A}$.

On the other hand, the {\it right unfold} ({\sc ru}) rule
selects some recursive atom $q$ and
performs an unfold on the RHS of a proof obligation wrt. $q$. 
In the proof process, the two unfold rules will
be systematically interleaved.

\begin{example}\label{ex:sub-ru}
Consider the following proof obligation:

\stuff{
$\tilde{A} \vdash \listn(x, L) \models \lseg(x, y, L_1), \listn(y, L_2), L \heq L_1 \hsep L_2$.
}

\end{example}


%
%

\begin{figure}

\begin{center}
${
\inferrule* [left=\sc (sub)]
  {\inferrule* [Left=\sc (ru)] 
    {\inferrule* [Left=\sc (cp)]
      {\texttt{True}}
      {\tilde{A} \vdash \listn(x, L) \models x = x, L_1 \heq \hemp, L \heq L_1 \hsep L}
    }
    {\tilde{A} \vdash \listn(x, L) \models \lseg(x, x, L_1), L \heq L_1 \hsep L}
  }
  {\tilde{A} \vdash \listn(x, L) \models \lseg(x, y, L_1), \listn(y, L_2), L \heq L_1 \hsep L_2}
}$
\end{center}

\caption{Proving with just U+M}
\label{prove-u+m}
\vspace{-2mm}
\end{figure}


\noindent In Fig.~\ref{prove-u+m}, we show how this proof obligation 
can be successfully dispensed by 
applying ({\sc sub}), ({\sc ru}), and ({\sc cp}) rules in sequence.
Note how  the ({\sc sub}) rule binds the existential variable $y$
to $x$ and simplifies the RHS of the proof obligation.


%





\noindent\textbullet~~The {\it induction applications}, namely {\sc (ia-1)} and {\sc (ia-2)} rules, 
transform the current obligation by making use of an assumption which has been added 
by the {\sc (lu+i)} rule. The two rules, also called the ``induction rules'' for short,
allow us to treat previously
encountered obligations as possible induction hypotheses.

Instead of directly proving the current obligation $\Left \models \Right$,
we now proceed by finding $\overline{\Left}$ and
$\overline{\Right}$ such that
$\Left \models \overline{\Left} \models \overline{\Right} \models \Right$.
The key here is to find those candidate goals where the validity of
$\overline{\Left} \models \overline{\Right}$ directly follows 
from a ``similar'' assumption $A$, together with $\theta$
to rename all the variables in $A$ to the variables in the current obligation, namely $\Left \models \Right$.
Assumption $A$ is an obligation which has been previously 
encountered in the proof process, and $A\theta$ assumed to be true,
as an induction hypothesis.
Particularly, we choose $\overline{\Left}$ and $\overline{\Right}$
so we can (easily) find  a renaming $\theta$ such that
$A\theta \implies \overline{\Left} \models \overline{\Right}$
($\implies$ denotes logical implication).

To be more deterministic and to prevent us from transforming 
to obligations harder than the original one, we require that
at least one of the remaining two entailments, namely 
$\Left \models \overline{\Left}$ and $\overline{\Right} \models \Right$, 
is discharged quickly by a direct proof.

In ({\sc ia-1}) rule, given the current obligation $p(\tilde{x}) \wedge \Left_1 \wedge \Left_2 \models \Right$
and an assumption $A \equiv p(\tilde{y}) \wedge \Left'   \models \Right'$,
we choose $p(\tilde{x}) \wedge \Left'\theta \wedge \Left_2$ to be our $\overline{\Left}$ 
and $\Right'\theta \wedge \Left_2$ to be our $\overline{\Right}$.
We can see that the validity of $\overline{\Left} \models \overline{\Right}$ directly follows 
from the assumption $A\theta$. 
One restriction onto the renaming $\theta$, to avoid circular reasoning, 
is that $\theta$ must rename $\tilde{y}$ to $\tilde{x}$ where
$p(\tilde{x})$ is an atom which has been generated
after $p(\tilde{y})$ had been unfolded. 
Such fact is indicated by ${\tt gen}(p(\tilde{x})) \geq {\tt kill}(p(\tilde{y}))$ in our rule.
While ${\tt gen}(p)$ denotes the timestamp when the recursive atom $p$ is generated
during the proof process, ${\tt kill}(p)$ denotes the timestamp
when $p$ is unfolded and removed. 
Another side condition for this rule 
is that the validity of $\Left \models \overline{\Left}$, or equivalently,
$\Left_1 \models \Left'\theta$ is discharged immediately by a direct proof.

In ({\sc ia-2}) rule,  given the current obligation $p(\tilde{x}) \wedge \Left_1 \models \Right$
 and an assumption $A \equiv p(\tilde{y}) \wedge \Left'   \models \Right'$,
 on the other hand, $p(\tilde{y})\theta \wedge \Left'\theta$ 
serves as our $\overline{\Left}$ while $\Right'\theta$ serves as our $\overline{\Right}$.
The validity of $\overline{\Left} \models \overline{\Right}$ trivially follows 
from the assumption $A\theta$, namely $p(\tilde{x}) \wedge \Left'\theta \models \Right'\theta$. 
As in ({\sc ia-1}), we also put similar restriction upon the renaming $\theta$.
Another side condition we require
is that  the validity of $\overline{\Right} \models \Right$ 
can be discharged immediately by a direct proof. 
At this point we could see the duality nature of ({\sc ia-1}) and ({\sc ia-2}).

Now let us briefly and intuitively explain the restriction upon the renaming $\theta$.
Here we make sure that $\theta$ renames atom $p(\tilde{y})$ to 
atom $p(\tilde{x})$, where  $p(\tilde{x})$ has been generated after 
$p(\tilde{y})$ had been unfolded (and removed). 
This helps to rule out certain potential $\theta$ which does not correspond 
to a number of left unfolds.
Such restriction helps ensure \emph{progressiveness} in the proof process
before the induction rules can take place. 
Otherwise, assuming the truth of $A\theta$ in constructing the proof for $A$ 
might not be valid.
This is the reason why for each element of $\tilde{A}$, we not only
keep track of the assumption, 
but also the recursive atom $p$ to which the application of 
({\sc lu+i}) gives rise to
the addition of  such assumption.

\ignore{  
We mention here that some works on recursive rules for defining
program properties, e.g. \cite{nguyen10shape3}, require some
well-founded metric $n$ associated to the predicate being defined so
that any recursive ``call'' would have to be associated with a metric
$m$ such that $m < n$.  (The non-recursive part of the definition would
have measure 0.)  With such a restriction, there is no danger of
circular reasoning when applied in our inductive framework.  
This is essentially because an unfolded obligation, when required to
behave like an obligation with metric $n$ before the unfolding, can only be associated
to another expression whose metric is less than $n$.
We however consider this kind of restriction on the rules as a significant
imposition.  
}

It is important to note that, our framework as it stands, does
not require any consideration of a base case, nor any well-founded
measure.  Instead, we depend on the Least Model Semantics (LMS) 
of our assertion language and the above-mentioned restrictions on the
renaming $\theta$.  In other words, we constrain the use of the rules,
which is transparent to the user, in order to achieve a well-founded 
conclusion.

Let us shed some insights with simple and concrete examples.
Consider the recursive predicate {\tt p}, defined as

\begin{center}
\stuff{
p($x$) \clpif~ p($x$).
}
\end{center}

\noindent
and the following two proof obligations:

\vspace{-3mm}
\numberwithin{equation}{section}
\begin{align}
{\tt p}(x) \models \listn(x, L) \label{ob:rules:1}\\
 \listn(x, L) \models {\tt p}(x) \label{ob:rules:2}
\end{align}

\noindent
We will now demonstrate that our method can 
prove \eqref{ob:rules:1}, but not \eqref{ob:rules:2}. 
We remark that \eqref{ob:rules:1} holds because 
under the Least Model Semantics (LMS) the LHS has no model; therefore no
refutation can be found regardless of what is
in the RHS~\footnote{i.e., {\em false} implies anything}.
On the other hand, \eqref{ob:rules:2} does not
hold because  $x = 0$ (and $L \heq \hemp$) is
a model of the LHS, but not a model of the RHS.

\begin{figure}[tbh]
\begin{center}
${
    {\inferrule* [Right=({\tt p}($x$))]
      {(\dagger)~ {\tt p}(x) \models \listn(x, L)}
      {(\dagger)~ {\tt p}(x) \models \listn(x, L)}
    }
}$
\end{center}
\vspace{-2mm}
\caption{Cyclic Proof for \eqref{ob:rules:1}}
\label{rule:prove-cyclic}
\end{figure}

\noindent
We start the discussion by first showing how ``Cyclic Proof'' would
handle \eqref{ob:rules:1}. See Fig.~\ref{rule:prove-cyclic}.
After unfolding the predicate {\tt p}($x$) in the LHS,
we immediately observe the same obligation, i.e., a
cyclic path has been established. Consequently, we can terminate the 
proof path and discharge the original obligation.

\begin{figure}[t]
\begin{center}
${
\inferrule* [left=\sc (lu+i)]
  {\inferrule* [Left=\sc (ia-1)] 
    {\inferrule* [Left=\sc (cp)]
      {\texttt{True}}
      {\{A\} \vdash \listn(x, L) \models \listn(x, L)}
    }
    {\{A\} \vdash {\tt p}(x) \models \listn(x, L)}
  }
  {\emptyset \vdash {\tt p}(x) \models \listn(x, L)}
}$
\end{center}
\vspace{-2mm}
\caption{Our Proof for \eqref{ob:rules:1}}
\label{rule:prove-induction}
\end{figure}
We now proceed showing in Fig.~\ref{rule:prove-induction} 
how our method would handle this obligation. We first perform a left unfolding,
adding $A \equiv \pair{{\tt p}(x) \models \listn(x, L)}{{\tt p}(x)}$
into the set of assumptions. Note that this unfolding step kills
the predicate {\tt p}($x$) and generates a new predicate {\tt p}($x$).
Thus the rule {\sc (ia-1)} is applicable now. We then re-write the 
LHS from {\tt p}($x$) to \listn($x, L$). Finally the proof succeeds
by consulting constraint solver, treating \listn($x, L$) as uninterpreted.

\begin{figure}[tbh]
\begin{center}
{$
  {\inferrule* [Left=\sc (ru)] 
    {\inferrule* [Left=\sc (lu+i)]
      {{\{A'\} \vdash x = 0, L \heq \hemp \models {\tt p}(x)}
      \\ \hspace{4mm}
       {\inferrule* [Left=,]{\vdots}{}}{}}
      {\emptyset \vdash \listn(x, L) \models {\tt p}(x)}
    }
    {\emptyset \vdash \listn(x, L) \models {\tt p}(x)}
  }
$}
\end{center}
\vspace{-2mm}
\caption{An Unsuccessful Attempt for \eqref{ob:rules:2}}
\label{rule:prove-induction-fail}
\end{figure}

\noindent
In contrast, now consider obligation \eqref{ob:rules:2} 
in Fig~\ref{rule:prove-induction-fail}. Obviously,
a direct proof for this is not successful. However,
if we proceed by a right unfold first, we get back the same obligation.
Different from before, and importantly, now no new assumption is added. 
We can see that the step does not help us progress
and therefore performing right unfold repetitively would get us nowhere.
Now consider performing a left unfold on the obligation.
The proof succeeds if we can discharge both

\begin{center} 
\stuff{
$\{A'\} \vdash x = 0, L \heq \hemp \models p(x)$ and \\
$\{A'\} \vdash L \heq \one{x}{t} \hsep L_1, \listn(t,L_1) \models p(x)$,
}
\end{center}

\noindent
where $A' \equiv \pair{\listn(x, L) \models {\tt p}(x)}{\listn(x, L)}$.

Focus on the obligation $\{A'\} \vdash x = 0, L \heq \hemp \models p(x)$.
Clearly consulting a constraint solver or performing substitution does not help.
Rule {\sc (lu+i)} is not applicable since no recursive predicate on the LHS.
As before, we cannot progress using {\sc (ru)} rule.
Importantly, the side conditions prevent {\sc (ia-1)} and {\sc (ia-2)}
from taking place. In summary, with our proof rules, 
this (wrong fact) cannot be established.

\subsection{Proving the Two Motivating Examples}
\label{rules:motivating}

\begin{figure*}[htb]
\small{
\begin{flushright} 
$
\inferrule* [Left=\scriptsize(LU+I)]
  {{\inferrule* [Left=\scriptsize (IA-1),rightskip=1em] 
    {\inferrule* [Left=\scriptsize (LU+I)]
      {{\inferrule* [Left=\scriptsize (RU),rightskip=1em]
        {\inferrule* [Left=\scriptsize (IA-1)]
          {\inferrule* [Left=\scriptsize (SUB)]
            {\inferrule* [Left=\scriptsize (CP)]
              {\texttt{True}}
               {\{A_1,A_2\} \vdash x \neq y, L \heq \one{x}{t} \hsep L_1, t \neq y, L_1 \heq \one{z}{y} \hsep L_2 \models x \neq y, L \heq \one{z}{y} \hsep \one{x}{t} \hsep L_2}
            }
            {\{A_1,A_2\} \vdash x \neq y, L \heq \one{x}{t} \hsep L_1, t \neq y, L_1 \heq \one{z}{y} \hsep L_2, \lseg(x, z, \one{x}{t} \hsep L_2) \models x \neq y, L \heq \one{z_1}{y} \hsep L_3, \lseg(x,z_1, L_3)}
          }
          {\{A_1,A_2\} \vdash x \neq y, L \heq \one{x}{t} \hsep L_1, t \neq y, L_1 \heq \one{z}{y} \hsep L_2, \lseg(t, z,L_2) \models x \neq y, L \heq \one{z_1}{y} \hsep L_3, \lseg(x,z_1, L_3)}
        }
        {\{A_1,A_2\} \vdash x \neq y, L \heq \one{x}{t} \hsep L_1, t \neq y, L_1 \heq \one{z}{y} \hsep L_2, \lseg(t, z,L_2) \models \lseg(x,y,L)}
      }  \\ {\inferrule* [Left=]{\vdots}{}}{}}
         {\{A_1\} \vdash  x \neq y, L \heq \one{x}{t} \hsep L_1, \lseg(t,y,L_1) \models \lseg(x,y,L)}
    }
    {\{A_1\} \vdash  x \neq y, L \heq \one{x}{t} \hsep L_1, \lsegleftmath(t,y,L_1) \models \lseg(x,y,L)}
  } \\ {\inferrule* [Left=,]{\vdots}{}}{}}
  {\emptyset \vdash \lsegleftmath(x,y,L) \models \lseg(x,y,L)}
$
\end{flushright}
\begin{center}
\textbf {where ${A_1 ~{\equiv}~ \pair{\lsegleftmath(x,y,L)  ~{\models}~ \lseg(x,y,L)}{\lsegleftmath(x,y,L)}}$ and
${A_2 ~{\equiv}~ \pair{x \neq y, L \heq \one{x}{t} \hsep L_1, \lseg(t,y,L_1) \models \lseg(x,y,L)}{\lseg(t,y,L_1)} }$}
\end{center}
}
\vspace{-2mm}
\caption{Proving \lsegleft$(x, y, L) ~{\models}~ \lseg(x, y, L)$.}
\label{fig:lsegproof}
\end{figure*}

\begin{figure*}[tbh] 
\small{
\begin{center}
${
\inferrule* [Left=\scriptsize (LU+I)]
  {
  {\inferrule* [Left=\scriptsize (IA-2)]
    {\inferrule* [Left=\scriptsize (RU)]
      {\inferrule* [Left=\scriptsize (SUB)]
        {\inferrule* [Left=\scriptsize (CP)]
          {\texttt{True}}
         {\{A\} \vdash  x \neq y, L_1 \heq \one{t}{y} \hsep L_3, \listn(y, L_2),  L_1 \hsep L_2 \models L_4 \heq \one{t}{y} \hsep L_2,  L_1 \hsep L_2 \heq L_3 \hsep L_4}
        }
        {\{A\} \vdash  x \neq y, L_1 \heq \one{t}{y} \hsep L_3, \listn(y, L_2),  L_1 \hsep L_2 \models L_4 \heq \one{t}{y_1} \hsep L_5, \listn(y_1, L_5), L_1 \hsep L_2 \heq L_3 \hsep L_4}
      }
      {\{A\} \vdash  x \neq y, L_1 \heq \one{t}{y} \hsep L_3, \listn(y, L_2),  L_1 \hsep L_2 \models \listn(t, L_4), L_1 \hsep L_2 \heq L_3 \hsep L_4}
    }
    {\{A\} \vdash  x \neq y, L_1 \heq \one{t}{y} \hsep L_3, \lseg(x,t,L_3),  \listn(y, L_2), L_1 \hsep L_2 \models \listn(x, L), L \heq L_1 \hsep L_2} 
  } \\ {\inferrule* [Left=,]{\vdots}{}}{}}
  {\emptyset \vdash \lseg(x,y,L_1), \listn(y, L_2), L_1 \hsep L_2 \models \listn(x, L), L \heq L_1 \hsep L_2}
}$\\
\vspace{2mm}
\textbf {where ${ A \equiv \pair{\lseg(x,y,L_1), \listn(y, L_2), L_1 \hsep L_2 \models \listn(x, L), L \heq L_1 \hsep L_2}{\lseg(x,y,L_1)} }$}
\end{center}
}
\vspace{-2mm}
\caption{Proving $\lseg(x,y,L_1), \listn(y, L_2), L_1 \hsep L_2 \models \listn(x, L), L \heq L_1 \hsep L_2$.}
\label{fig:llproof}
\end{figure*}

Let us now revisit the two motivating examples introduced earlier, 
on which both U+M and ``Cyclic Proof'' are not effective. The main reason
is that both examples involve unmatchable predicates while at the same time
exhibiting ``recursion divergence''.

\begin{example}\label{ex:induction1}
Consider the entailment relating two definitions of list segments:
\lsegleft$(x,y,L) \models \lseg(x,y,L).$
\end{example}

\noindent
%
Our method can discharge this obligation by applying  ({\sc ia-1}) rule twice.
For space reason, in Fig.~\ref{fig:lsegproof}, we only show the 
interesting path of the proof tree (leftmost position).
First, we unfold the predicate $\lsegleftmath(x,y,L)$ 
in the LHS of the given obligation via ({\sc lu+i}) rule.
The original obligation, while being removed, is added as an assumption 
$A_1$. We next make use of $A_1$ as an induction hypothesis 
to perform a re-writing step, i.e., an application of ({\sc ia-1}) rule.
Similarly, in the third step, we unfold the predicate $\lseg(t,y,L_1)$ in the LHS 
via ({\sc lu+i}) rule and add the assumption $A_2$.
After unfolding in the RHS via ({\sc ru}) rule and re-writing 
with the induction hypothesis $A_2$
using ({\sc ia-1}) rule, we are able to bind the existential variable $z_1$ to $z$ and
simplify both sides of the proof obligation using ({\sc sub}) rule.
Finally, the proof path is terminated by consulting a
constraint solver, i.e., ({\sc cp}) rule.


\begin{example}\label{ex:induction2}
Consider the entailment: 

$\lseg(x,y,L_1), \listn(y, L_2), L_1 \hsep L_2 \models \listn(x, L), L \heq L_1 \hsep L_2$.
\end{example}
 
\noindent
Fig.~\ref{fig:llproof} shows, only the interesting proof path, 
how we can successfully prove this entailment using the ({\sc ia-2}) rule.
We first unfold $\lseg(x,y,L_1)$ in the LHS, adding $A$
into the set of assumptions.
Then using $A$ as an induction hypothesis, we can rewrite the current 
obligation via ({\sc ia-2}) rule. 
Note that, here we use ({\sc ia-2}) rule instead of ({\sc ia-1}) 
rule as in previous example.
After applying ({\sc ru}) rule, we are able to bind the existential variable $y_1$ to $y$ and
simplify both sides of the proof obligation with ({\sc sub}) rule.
Finally, the proof path is terminated by consulting a
constraint solver, i.e., ({\sc cp}) rule.


Let us pay a closer attention at the step where we 
attempt re-writing, making using the available induction hypothesis.
For the sake of discussion, assume that instead of ({\sc ia-2}) we now
attempt to apply rule ({\sc ia-1}). The requirement for $\theta$
forces it to rename $x$ to $x$ and $y$ to $t$.
However, the side condition $\Left_1 \models_{\tt DP} \Left'\theta$
cannot be fulfilled, since $x \neq y, L_1 \heq \one{t}{y} \hsep L_3, 
\listn(y, L_2), L_1 \hsep L_2 \not\models_{\tt DP}  \listn(t,\_)$.

Now return to the attempt of ({\sc ia-2}) rule.
The RHS of the current obligation matches with the RHS of the \emph{only}
induction hypothesis perfectly. This matching requires $\theta$ to rename $x$ back to $x$.
On the LHS, we further require $\theta$ to rename $y$ to $t$ so that
$\lseg(x,t) \equiv \lseg(x,y)\theta$. Note that
$\lseg(x,t)$ was indeed generated after $\lseg(x,y)$ had been unfolded and removed
(i.e., killed). The remaining transformation is straightforward.

\ignore{

The process of using induction is akin to the proof of equations defined by
term-rewriting systems, where the folding/unfolding serves to
arrive at a {\em normal form} so that equality reduces to
the problem of having the same normal form\footnote{
Strictly speaking, we also require that the system
is {\em confluent}.}.
}


\subsection{Soundness}

%

\begin{theorem}[Soundness]
An entailment $\Left \models \Right$ holds if,
starting with $\emptyset \vdash \Left \models \Right$,
there exists a sequence of applications of proof rules that
results in an empty set of proof obligations.
\qed
\end{theorem}


\noindent
\textbf{Proof Sketch.}
The soundness of rule \textsc{(cp)} is obvious.                                                      
The rule \textsc{(ru)} is sound because when
$\Right' \in \unfold_{q}(\Right)$ then $\Right' \models \Right.$ Therefore, the proof
of $\tilde{A} \vdash \Left \models \Right$ can be replaced by
the proof of $\tilde{A} \vdash \Left \models \Right'$ since
$\Left \models \Right'$ is stronger than $\Left \models \Right.$ 
Similarly, the rule \textsc{(sub)} is sound because $\Left \wedge p(\tilde{x}) \models \Right \theta$
and $\Left \wedge p(\tilde{x}) \models_{\tt CP} \tilde{x} = \tilde{y} \theta$                                                is stronger than the $\Left \wedge p(\tilde{x})                                          
\models \Right \wedge p(\tilde{y}).$

The rule \textsc{(lu+i)} is {\em partially sound\/} in the sense that
when $\unfold_p(\Left) = \{\Left_1,\ldots,\Left_n\},$ then proving $\Left \models                                        
\Right$ can be substituted by proving $\Left_1 \models \Right, \ldots, \Left_n                                         
\models \Right.$ This is because in the least-model semantics 
of the definitions, $\Left$ is equivalent to $\Left_1\vee\ldots\vee \Left_n.$  
However, whether the addition of $\pair{\Left \models \Right}{p}$
to the set of assumed obligations
$\tilde{A}$ is sound depends on the use of 
them in the application of \textsc{(ia-1)} and \textsc{(ia-2)}.

We now proceed to prove the soundness of \textsc{(ia-1)} and \textsc{(ia-2)}.
First, define a {\em refutation\/} to an obligation $\Left \models \Right$ as
a successful derivation of one or more atoms in $\Left$ whose answer
$\Psi$ has an instance (ground substitution) $\beta$ such that
$\Psi\beta \wedge \Right\beta$ is false. A finite refutation
corresponds to a such derivation of finite length. A nonexistence of
finite refutation means that $\lm(P) \models  (\Left \rightarrow \Right)$, or in other words,
$\Left \models \Right$. A
derivation of a refutation is obtainable by left unfold \textsc{(lu+i))}
rule only.  Hence a finite refutation of length $k$ implies a
corresponding $k$ left unfold \textsc{(lu+i)} applications that
results in a contradiction. 

In the rules \textsc{(ia-1)} and \textsc{(ia-2)},
we assume the hypothesis $A\theta$, where $A \equiv \Left' \models \Right'$
is some entailment encountered previously.
By having the side conditions proved separately, we then can soundly transform
the current entailment $B$ into a new entailment $C$. 
In case of \textsc{(ia-1)},
$B \equiv p(\tilde{x}) \wedge \Left_1 \wedge \Left_2 \models \Right$ and 
$C \equiv \Right'\theta \wedge \Left_2 \models \Right$.
 In case of \textsc{(ia-2)}, we have
$B \equiv p(\tilde{x}) \wedge \Left_1 \models \Right$ and 
$C \equiv \Left_1 \models \Left'\theta$.

Notice that the side conditions ensure that $A\theta \implies (C \implies B)$,
where $\implies$ denotes implication.
The side conditions also enforce the renaming $\theta$ to ``progress''
at least the left unfold of recursive atom $p(\tilde{y})$ to match
with a newly generated atom $p(\tilde{x})$. This indeed enforces 
a well-founded measure on $A$.

To be more concrete, note that our transformation from $B$ to $C$ is {\em unsound 
only if} there exists a refutation $\beta$ to $B$, and therefore $A$, but $\beta$
is not a refutation to $C$. 
We then proceed to prove by contradiction. W.l.o.g., assume $\beta$ is such a
refutation and is the refutation to $A$ which has the smallest length $k$.
Trivially $k > 0$ as $A$ has no finite refutation of length $0$.
Since there is at least one left unfold from $A$ to $B$, $\beta$
must be a refutation to $B$ but of length less than equal to $k$.
However, since $A\theta \implies (C \implies B)$, and $\beta$ is a refutation
of $B$ but not $C$, therefore $\beta$ is also a refutation of $A\theta$.
Since $\theta$ must ``progress'' $A$ by at least one left unfold, we end up with the fact that
$A$ has a refutation of length less than $k$. This is a contradiction.
\qed

\section{Implementation}
\label{sec:implementation}
\newcommand{\unioneq}{{_\cup}{=}}

\begin{figure}[bh!]
\begin{center}
\small{
$\begin{array}{rl}

\multicolumn{2}{l}{\mbox{\bf function} ~\func{Prove}(\Left \models \Right, ~\tilde{A}, ~lb, ~rb, ~ib)}  \\

     & \mbox{/* Natural proof, i.e. by unification and \smt{} */} \\
\ppp & \If ~(\func{DirectProof}(\Left \models \Right)) ~ \Return ~\true \\
\\

     & {\bf let} ~\Left = \Phi, \LeftD_1, \ldots, \LeftD_n ~and~ \Right = \Psi, \RightD_1, \ldots, \RightD_m \\
\ppp & OrSet ~\Assign ~\emptyset \\

\ppp & \If ~(ib < {\sf INDUCTIONBOUND}) \ptab \ptab \mbox{/* Induction Application */} \\
\ppp     & \ptab \Foreach (\pair{p(\tilde{y}) \wedge \Left' \models \Right')}{p(\tilde{y})} \in \tilde{A}) \\
\ppp     & \ptab \ptab \func{Find} ~ p(\tilde{x}) \in \Left \mbox { s.t. } {\tt gen}(p(\tilde{x})) \geq {\tt kill}(p(\tilde{y})) \\
\ppp     & \ptab \ptab \func{Find} ~ \Left_1 \subseteq \Left \mbox{ s.t. } \theta ~\Assign ~\func{DirectProof}(\Left_1 \models \Left') ~\neq ~\perp \\
\ppp 	   & \ptab \ptab \If (\tilde{x} = \tilde{y}\theta ~and~\theta~ \mbox{is a valid renaming}) \\
\ppp 	   & \ptab \ptab \ptab \Left_{new} \Assign  \Left \setminus \{p(\tilde{x})\} \setminus  \Left_1 \cup \Right'\theta \\
\ppp     & \ptab \ptab \ptab OrSet ~\unioneq \{\afivetuple{\Left_{new} \models \Right}{\tilde{A}}{lb}{rb}{ib+1}\} \\
\ppp     & \ptab \Foreach (\pair{p(\tilde{y}) \wedge \Left' \models \Right')}{p(\tilde{y})} \in \tilde{A})\\
\ppp     & \ptab \ptab \If ~(\theta_1 ~\Assign ~\func{DirectProof}(\Right' \models \Right) = ~\perp) ~{\bf continue} \\
\ppp     & \ptab \ptab \func{Find} ~ p(\tilde{x}) \in \Left \mbox { s.t. } {\tt gen}(p(\tilde{x})) \geq {\tt kill}(p(\tilde{y})) \\
\ppp     & \ptab \ptab \func{Find} ~\mbox {a valid renaming } 
\theta \supseteq \theta_1 \mbox{ s.t. } \tilde{x} = \tilde{y}\theta \\
\ppp     & \ptab \ptab OrSet ~\unioneq \{\afivetuple{\Left \setminus \{p(\tilde{x})\} \models \Left'\theta}{\tilde{A}}{lb}{rb}{ib+1}\} \\
\\

\ppp & \If ~(lb < {\sf MAXLEFTBOUND}) \ptab \ptab \mbox{/* Left Unfold */} \\

\ppp & \ptab \Foreach ~(\LeftD_i \in \Left) \\
\ppp & \ptab \ptab Obs ~\Assign  ~\emptyset \\
\ppp & \ptab \ptab \tilde{A}' ~\Assign ~\tilde{A} ~\cup ~\{\pair{\Left \models \Right}{\LeftD_i}\} \\
\ppp & \ptab \ptab \Foreach ~ (\Left_{j} \in (\{\Left_{1}, \Left_{2}, \ldots \Left_{l}\} ~\Assign ~{\sf UNFOLD}(\LeftD_i)))  \\

\ppp & \ptab \ptab \ptab ob ~\Assign 
    ~\afivetuple{(\Left_{j} \cup \Left \setminus \{\LeftD_i\}) \models \Right}{\tilde{A}'}{lb+1}{rb}{ib}\\
\ppp & \ptab \ptab \ptab \If ~ (\func{trivially\_true}(ob)) ~{\bf continue} \\
\ppp & \ptab \ptab \ptab Obs ~\Assign ~ Obs \cup \{ob\} \\
\ppp & \ptab \ptab \If ~(Obs = ~\emptyset) ~\Return ~\true ~\Else ~OrSet ~\unioneq ~\{Obs\} \\

\\
\ppp & \If ~(rb < {\sf MAXRIGHTBOUND} ~and ~ \neg \func{contradict}(\Left \models \Right)) \\
& \ptab \ptab  \mbox{/* Right Unfold */} \\

\ppp & \ptab \Foreach ~(\RightD_i \in \Right) \\
\ppp & \ptab \ptab \Foreach (~\Right_{j} \in \{\Right_{1}, \Right_{2}, \ldots \Right_{k}\} ~\Assign ~{\sf UNFOLD}(\RightD_i)) \\
\ppp & \ptab \ptab \ptab ob = \afivetuple{\Left \models (\Right_{j} \cup \Right \setminus \{\RightD_i\})}{\tilde{A}}{lb}{rb+1}{ib} \\
\ppp & \ptab \ptab \ptab OrSet ~\unioneq \{\{ob\}\} \\
\\
\ppp & \If ~(OrSet ~= \emptyset) ~\Return ~\false \\
\ppp \label{algo:heuristics} & OrSet ~\Assign ~\func{OrderByHeuristics}(OrSet) \\
\ppp & \Foreach ~(Obs \in OrSet) \\
\ppp & \ptab \If ~(\func{ProveAll}(Obs)) ~\Return ~\true \\
\ppp & \Return ~\false \\
\multicolumn{2}{l}{\mbox{\bf endfunction}}

\end{array}$
}
\vspace{-3mm}
\end{center}
\caption{The Main Algorithm\label{algm}}
\end{figure}

\begin{figure}[tbh]
\begin{center}
\small{
$\begin{array}{rl}

\multicolumn{2}{l}{\mbox{\bf function} ~\func{ProveAll}(Obs)}  \\
\ppp & \Foreach ~(\afivetuple{\Left \models \Right}{\tilde{A}}{lb}{rb}{ib} \in Obs) \\
\ppp & \ptab \If  ~(\neg ~\func{Prove}(\Left \models \Right, ~\tilde{A}, ~lb, ~rb, ~ib)) ~\Return ~\false; \\
\multicolumn{2}{l}{\mbox{\bf endfunction}} \\ 
\\

\multicolumn{2}{l}{\mbox{\bf function} ~\func{DirectProof}(\Left \models \Right)} \\
\ppp & \If ~(\exists ~q(\tilde{x}) \in \Right ~\mbox{such that} \not\exists ~q(\tilde{y}) \in \Left) 
          ~\Return ~ \perp \\
\ppp & \Left' ~\Assign  ~\func{get\_all\_recursive}(\Left) \\
\ppp & \Right' ~\Assign  ~\func{get\_all\_recursive}(\Right) \\
\ppp & \Theta ~\Assign ~\{\mbox{substitution}~ \theta ~|~ \Right'\theta \subseteq \Left' \} \\
\ppp & \If ~(\Theta ~= ~\emptyset) ~\Return ~\perp \\

\ppp & \Foreach ~(\theta  \in \Theta) \\
\ppp & \ptab \Phi ~\Assign ~\func{get\_all\_nonrecursive}(\Left) \\
\ppp & \ptab \Psi ~\Assign ~\func{get\_all\_nonrecursive}(\Right) \\
\ppp & \ptab \theta' ~\Assign ~\func{bind\_remaining\_existential\_variables}(\Psi, \Phi, \theta) \\
     & \ptab \mbox{/* Extend $\theta$ to $\theta'$ by trying obvious bindings} \\
     & \ptab \ptab \mbox{for remaining existential variables */}\\
\ppp & \ptab \If ~(\func{has\_existential\_variables}(\Psi, \Phi, \theta')) ~\bf{continue} \\
\ppp & \ptab \If ~(\func{entailment}(\Phi, \Psi \theta')) ~\Return ~\theta' \\
\ppp & \Return ~\perp\\
\multicolumn{2}{l}{\mbox{\bf endfunction}}\\

%
%
%
%
\end{array}$
}
\vspace{-3mm}
\end{center}
\caption{Supporting Functions\label{algm}}
\end{figure}

\noindent
Let us briefly highlight our implementation, which  
intuitively follows from the proof rules in Sec.~\ref{sec:proofrules}. 
The main algorithm is in Figure~\ref{algm}.
In the figure, we use $X~\unioneq~Y$ to denote
$X ~\Assign ~X \cup Y$.
 
We start off by calling the function $\func{Prove}$ with the 
original obligation $\Left \models \Right$, the set of assumptions
$\tilde{A}$ to be $\emptyset$, and all the counters $lb, rb, ib$ to be $0$.
The counters $lb, rb, ib$ are to keep track of, respectively,
how many left unfolds, right unfolds, and inductions have been
applied in this current path. These counters are to ensure 
that our algorithm terminates. In our experiments, the typical values for
{\tt INDUCTIONBOUND, MAXLEFTBOUND, MAXRIGHTBOUND}
are respectively $3$, $5$, $5$.

Typically an unoptimized proof obligation usually can be partitioned into
a number of smaller and simpler  proof obligations
(e.g., by eliminating redundant 
terms and variables). This step
can be implemented using any standard proof slicing technique
and is not the focus of our discussion.

\vspace{2mm}
\noindent
{\bf Base Case:}
The function \func{DirectProof}
acts as the base case of our recursive algorithm.
For each proof obligation, we first attempt a \emph{direct proof}, 
i.e., to discharge by 
applying rule ({\sc sub}) repetitively and then querying 
Z3 solver~\cite{z3}, after treating all recursive predicates in the 
LHS as uninterpreted, as in ({\sc cp})-rule.

Intuitively, this step succeeds if the proof obligation
is simple ``enough'' such that a {\it proof by matching} can be achieved.
We note here that, our proof rules in Section~\ref{sec:proofrules} allow
other rules, e.g., ({\sc ru}) rule in Example~\ref{ex:sub-ru}, to interleave
with the ({\sc sub}) and ({\sc cp}) rules.
However, in our deterministic implementation,
applications of ({\sc sub}) and ({\sc cp}) rules
are coupled together.

\ignore{
This function corresponds to the repetitive applications of the ({\sc sub})
rule and a successful query to an \smt{} solver, after treating 
all recursive predicates in the LHS as uninterpreted (({\sc cp})-rule).

Intuitively, \func{DirectProof} succeeds if the proof obligation
is simple ``enough'' such that a {\it proof by matching} can be achieved.
We note here that, our proof rules in Sec.~\ref{sec:proofrules} allow
other rules (e.g. ({\sc ru}) rule in Example~\ref{ex:sub-ru}) to interleave
with the ({\sc sub}) and ({\sc cp}) rules.
However, in our deterministic algorithm,
applications of ({\sc sub}) and ({\sc cp}) rules
are coupled together within the function \func{DirectProof}.
}

\ignore{
Our function \func{DirectProof} is similar to the process 
of performing formula abstraction and 
then submitting the final formula to an \smt{} solver. 
The main difference is that, in their restricted setting, finding the appropriate
substitution for existential variables in the RHS is ``deterministic''.
Here the task requires some amount of search. However,
the cost of this search is negligible, since the number of 
recursive terms appear in the proof obligation is small.
}

Let us examine the function \func{DirectProof}.
If there is a recursive predicate $q$ on the RHS, but not in the LHS,
the function returns immediately, indicating failure with $\perp$.
Otherwise, the function then proceeds by finding some (not exhaustive)
substitutions $\Theta$ such that with each $\theta \in \Theta$,
we can simultaneously remove all the recursive predicates on the
RHS. This process will remove most existential variables on the RHS,
since existential variables usually appear in some recursive predicates.

In case there remain some existential variables on the RHS,
we attempt to bind them with the obvious candidates on the LHS
(therefore extend $\theta$ to $\theta'$).
%
%
%
%
%
%
After this attempt, if the RHS contains no existential variables,
we then call an \smt{} solver for entailment check.
If the answer is yes, $\theta'$ is returned,
indicating that a direct proof has been achieved.

\vspace{2mm}
\noindent
{\bf Recursive Call:}
When the attempt of direct proof is not successful,
we collect all possible transformations of the current proof obligation,
using  ({\sc ia-1}), ({\sc ia-2}), ({\sc lu+i}),
({\sc ru}) rules, into \emph{a set of set of} obligations $OrSet$. 
The current proof obligation can be successfully discharged
if there is {\em any} set of proof obligations $Obs \in OrSet$, where
we can discharge {\em every} proof obligation $ob \in Obs$.
The realization of the proof rules in our algorithm is straightforward, except
for a few noteworthy points:

\begin{enumerate}
\item
Our induction applications will not exhaustively
search for all possible candidates. Instead, we only search
for some trivial renaming which meets the side conditions of the rules.

\item When we perform left unfold, an obligation which is trivially true
(\func{trivially\_true}),
i.e. the non-recursive part of the LHS is unsatisfiable,  
is immediately removed.

\item If the current obligation contains the LHS and RHS which
contradict each other (\func{contradict}), right unfold will be avoided. 
The proof for this obligation can succeed only if 
there are no models for the LHS 
(so only left unfolds are required).

\end{enumerate}

\noindent 
We note that our proof search proceeds recursively in a depth first search manner.
The order in which the sets of obligations $Obs \in OrSet$ are considered
might heavily affect the efficiency, i.e. the running time, 
but {\em not} the {\em effectiveness}, i.e. the ability to prove, of our framework.
Such order is dictated by our heuristics, as the call
to function \func{OrderByHeuristics} (line~\ref{algo:heuristics}) indicates.
We remark that our heuristics, described below, are very {\em intuitive}
and directly follow from the fact that
our base case is reached  by a successful direct proof.



We proceed by a number of passes. 
In each pass, we first order the obligations within each $Obs \in OrderSet$.
We then consider the order of $OrderSet$ by comparing
the last obligation in each set $Obs \in OrderSet$.
Subsequent passes will not
undo the work of the previous passes, but instead 
work on the obligations and/or sets of obligations
which are tied in previous passes.

\begin{enumerate}
\item
An obligation which has contradicting LHS and RHS, given by the function $\func{contradict}$
will be ordered after those do not (since the chance to successfully
discharge such obligation is small).

\item
An obligation contains no recursive predicates on the RHS 
will be order before those contain some recursive predicate(s) on the RHS.

\item
An obligation having a recursive predicate $q$ such that $q$ appears
in the RHS but not in the LHS will be ordered after those not.

\item
An obligation contains more existential variables which cannot be
deterministically bound to some non-existential variables (variables on the LHS) will
be ordered after those contains less.

\item 
An obligation resulted from a left unfold will be ordered before 
those resulted from a right unfold (since it allows 
an induction hypothesis to be added).

\end{enumerate}

\begin{example}
Revisit the obligation in Example~\ref{ex:sub-ru}, but now
with the starting set of assumptions to be empty:

\begin{center}
\stuff{
$\emptyset \vdash \listn(x,L) \models \lseg(x,y,L_1), \listn(y,L_2), L \heq L_1 \hsep L_2$.
}
\end{center} 
\end{example}

\noindent
For simplicity we ignore the information about the counters
$lb, rb,$ and $ib$. 
First, the call to \func{DirectProof} fails since there is
the predicate $\lseg$ which appears in the RHS but not in
the LHS. Induction rules cannot take place either, as the set
of assumptions is currently empty.
We prooceed by performing a left unfold first. Note that there is only one recursive predicate on the LHS.
Let $\tilde{A}$ be:

\begin{small}
\begin{center}
$\{\pair{\listn(x,L) \models \lseg(x,y,L_1), \listn(y,L_2), L \heq L_1 \hsep L_2}{\listn(x,L)}\}$.
\end{center}
\end{small}

\noindent
The result for our left unfold is a set of two obligations:

\begin{center}
\small{
\stuff{
$O_0 \equiv \{ \tilde{A}' \vdash {x}={0}, L \heq \hemp \models \lseg(x,y,L_1), \listn(y,L_2),  L \heq L_1 \hsep L_2$; \\
$\tilde{A} \vdash L {\heq} ({x}\mapsto{t}) \hsep L', \listn(t,L') \models \lseg(x,y,L_1), \listn(y,L_2), L{\heq}L_1 \hsep L_2\}$ 
}
}
\end{center}

\noindent
We proceed with right unfold, producing four sets, each consists of one (simplified) obligation as follows:

\begin{center}
\small{
\stuff{
$O_1 \equiv \{\emptyset \vdash \listn(x,L) \models \lseg(x,y,L_1), y = 0, L_2 \heq \hemp, L {\heq} L_1 \hsep L_2\}$ \\
$O_2 \equiv \{\emptyset \vdash \listn(x,L) \models \lseg(x,y,L_1),  \listn(t,L_3), L {\heq} L_1 \hsep ({y}\mapsto{t}) \hsep L_3 \}$ \\
$O_3 \equiv \{\emptyset \vdash \listn(x,L) \models {x}={y}, \listn(y, L_2), L \heq \hemp \hsep L_2\}$ \\
$ O_4 \equiv\{\emptyset \vdash \listn(x,L) \models x {\neq} y, \lseg(x,t,L_3), \listn(y,L_2),  L {\heq} ({t}\mapsto{y}) \hsep L_3 \hsep L_2\}$
}
}
\end{center}

\noindent
Assume that the initial order of those sets of obligations are as shown above.
After the first two passes, the order between those sets is the same.
The third pass, however, moves the singleton set $O_3$
to the first position. The fourth pass, on the other hand, moves 
$O_1$ to the second position.
The fifth pass keep $O_0$
at the third position. The remaining two  singleton sets,
namely $O_2$ and $O_4$
are tied and placed at the end.

We proceed with the first obligation set, namely $O_3 $,
and a direct proof of it is successful. Therefore the original 
obligation can be discharged.
The corresponding sequence  of the proof rules 
is shown below, which is slightly different from what shown 
in 
Fig.~\ref{prove-u+m}. 

\begin{center}
${
\inferrule* [left=\sc (ru)]
  {\inferrule* [Left=\sc (sub)] 
    {\inferrule* [Left=\sc (cp)]
      {\texttt{True}}
      {\emptyset \vdash \listn(x,L) \models x = x, L \heq \hemp \hsep L}
    }
    {\emptyset \vdash \listn(x,L) \models x = y, \listn(y,L_2), L \heq \hemp \hsep L_2}
  }
  {\emptyset \vdash \listn(x,L) \models \lseg(x,y,L_1), \listn(y,L_2), L \heq L_1 \hsep L_2 }
}$
\end{center}




%

%



\section{Experiments}
\label{sec:experiment}

Our evaluations are performed on a 3.2Gz Intel processor with 2GB RAM, running Linux. 
We evaluated our prototype on a comprehensive set of benchmarks, 
including both academic algorithms and real programs.
The benchmarks are collected from existing 
systems~\cite{nguyen08cav,nguyen10shape3,madhusudan12dryad,qiu13dryad,cyclic12aplas},
those considered as the state-of-the-art for the purpose of
proving user-defined recursive data-structure properties in imperative languages.
We first demonstrate our evaluation with benchmarks that the state-of-the-art can handle,
then with ones that are beyond their current supports.

\subsection{Within the State-of-the-art}

In this subsection, we consider the set of proof obligations where the
state-of-the-art, i.e., U+M and ``Cyclic Proof'', are effective.
Given that our proof method theoretically subsumes the state-of-the-art,
the main purpose of this study is to evaluate the \emph{efficiency} 
of our implementation against existing systems.
To some extent, this exercise serves as a sanity 
check for our implementation.

We first start with proof obligations where U+M can automatically
discharge without the help of user-defined lemmas. 
They are collected from the benchmarks of U+M frameworks
\cite{nguyen10shape3,madhusudan12dryad,qiu13dryad}.
As expected, our prototype prove all of those obligations,
the running time for each is negligible (less than 0.2 second).
This is because the proof obligations usually require just either
\emph{one} left unfold or \emph{one} right unfold 
before matching --- i.e., a direct proof --- can successfully take place.

The second set of benchmarks are from ``Cyclic Proof'' \cite{cyclic12aplas},
which are also used in {\sf SMT-COMP 2014} (Separation Logic)\footnote{See
https://github.com/mihasighi/smtcomp14-sl}.
They are proof obligations which involve unmatchable predicates, 
thus U+M will not be effective.
We also succeed in proving all of those obligations, 
less than a second for each.

In summary, the results demonstrate that 
(1) our prototype is able to
handle what the state-of-the-art can;
(2) our implementation is \emph{competitive} enough.

\subsection{Beyond the State-of-the-art}

We now move to demonstrate the \emph{key} result of
this paper: proving what are beyond the state-of-the-art.

\vspace{1mm}
\noindent
{\bf Proving User-Defined Lemmas:}
Our prototype can prove all commonly used lemmas, collected from 
\cite{nguyen08cav,nguyen10shape3,madhusudan12dryad,qiu13dryad},
which U+M and ``Cyclic Proof'' cannot handle.
The running time is always less than a second for each lemma.
Table~\ref{table:lemma} shows a non-exhaustive list of
common user-defined lemmas.
We purposely abstract them from the original usage in order 
to make them general and representative enough.
The lemmas are written in traditional Separation Logic syntax 
for succinctness. Note that due to the duality of the definitions for
list segments, e.g., \lseg ~vs.~\lsegleft,
each lemma containing them would usually has a dual version, 
which for space reason we do not list down in Table~\ref{table:lemma}.
Similarly, some extensions, e.g.,  to capture the relationship of collective 
data values (using sets or sequences) between the LHS and the RHS,
while can be automatically discharged by our prototype,
are not listed in the table.

\begin{table}[tbh]
\begin{center}
\begin{small}
\begin{tabular}{|l|l|}
\hline
\textsf{~~~~~~~~~~~~~~~~~~~~~~~~~~~~~~~~~~~~Lemma}\\
\hline
\hline

\rule{0pt}{2.5ex}
\sortedlist($x,min$) $\models$ $\listn$($x$)\\
\rule{0pt}{2.5ex}

\sortedlist$_1$($x,len,min$) $\models$ $\listn_1$($x,len$)\\
\rule{0pt}{2.5ex}

\sortedlist$_1$($x,len,min$) $\models$ \sortedlist($x,min$)\\
\rule{0pt}{2.5ex}


\sortedls($x,y,min,max$) $\hsep$ \sortedlist($y,min_2$) $\wedge$ $max \leq min_2$ $\models$\\
$\> \> \> \> \> $\sortedlist($x,min$)\\

\hline

\rule{0pt}{2.5ex}
\lseg($x,y$) $\hsep$ \listn($y$) $\models$ \listn($x$)\\
\rule{0pt}{2.5ex}

\lseg($x,y$) $\models$ \lsegleft($x,y$) ~and~ \lsegleft($x,y$) $\models$ \lseg($x,y$)\\
\rule{0pt}{2.5ex}

\lsegleft$_1$($x,y,len_1$) $\hsep$ \lsegleft$_1$($y,z,len_2$) $\models$ \lsegleft$_1$($x,z,len_1{+}len_2$)\\
\rule{0pt}{2.5ex}

\lseg$_1$($x,y,len_1$) $\hsep$ \listn$_1$($y,len_2$) $\models$ \listn$_1$($x,len_1{+}len_2$)\\
\rule{0pt}{2.5ex}

\lsegleft$_1$($x,last,len$) $\hsep$ ($last \mapsto new$) $\models$ \lsegleft$_1$($x,new,len+1$)\\

\hline

\rule{0pt}{2.5ex}
\dlseg($x,y$) $\hsep$ \dlistn($y$) $\models$ \dlistn($x$)\\
\rule{0pt}{2.5ex}

\dlsegleft$_1$($x,y,len_1$) $\hsep$ \dlsegleft$_1$($y,z,len_2$) $\models$ \dlsegleft$_1$($x,z,len_1{+}len_2$)\\
\rule{0pt}{2.5ex}

\dlseg$_1$($x,y,len_1$) $\hsep$ \dlistn$_1$($y,len_2$) $\models$ \dlistn$_1$($x,len_1{+}len_2$)\\

\hline

\rule{0pt}{2.5ex}
{\sf avl}($x,height,min,max,balance$) $\models$ {\sf bstree}($x,height,min,max$)\\

\rule{0pt}{2.5ex}
{\sf bstree}($x,height,min,max$) $\models$ {\sf bintree}($x,height$)\\

\hline
\end{tabular}
\end{small}
\end{center}
\vspace{-8pt}
\caption{Proving lemmas (existing systems cannot prove).}
\label{table:lemma}
\end{table}

Let us briefly discuss Table~\ref{table:lemma}. 
The first group talks about sorted linked lists.
As an example, the second lemma is to state that a sorted list with
length $len$ and the minimum element $min$ is also a list with the same length.
We use indexes for different definitions of data structures,
which involve different properties (i.e., \sortedlist~ and \sortedlist$_1$).
The second, third and fourth groups are related to singly-linked lists,
doubly-linked lists, and trees respectively.



\vspace{1mm}
\noindent
{\bf Verifying Programs without Using Lemmas:}
Lemmas can serve many purposes.
One important usage of lemmas in U+M systems
is to equip a proof system with the power
of user-provided re-writing rules, so as to overcome
the main limitation of unfold-and-match.
However, in the context of program verification,
eliminating the usage of lemmas is crucial for improving the performance.
This is because lemma applications, coupled with unfolding, 
often induce very large search space.

We now use the set of academic algorithms and open-source library 
programs\footnote{See {\tt http://www.cs.uiuc.edu/$\sim$madhu/dryad/sl}}, 
collected and published by \cite{nguyen10shape3,qiu13dryad}, 
to demonstrate that our prototype can verify all of the programs in this set
without using lemmas.
The library programs include Glib open source library, the OpenBSD library, 
the Linux kernel, the memory regions and the page cache implementations 
from two different operating systems.
While Table~\ref{table:academic} summarizes the verification of 
data structures from academic algorithms, 
Table~\ref{table:real} reports on open-source library programs.

\begin{table}[tbh]

\begin{center}
\begin{small}
\begin{tabular}{|l|l|c|}
\hline
\textsf{Data-structure} & \textsf{Function}  
& \textsf{T/F}\\
\hline
\hline

\stuff{Sorted List} &
\stuff{find\_last\_iter, insert\_iter,\\
quick\_sort\_iter, bubble\_sort}
&
${<} 1s$ \\

\hline

\stuff{Circular List} &
\stuff{count}
&
${<} 1s$ \\


\hline

\stuff{Binary Search\\~~~~Tree} &
\stuff{insert\_iter,find\_leftmost\_iter,\\
remove\_root\_iter, delete\_iter}
&
${<} 1s$ \\

\hline
\end{tabular}
\end{small}
\end{center}
\vspace{-8pt}
\caption{Verification of Academic Algorithms (existing systems require lemmas).}
\label{table:academic}
\end{table}

\noindent \underline{\emph{Remark \#1:}} 
Using automatic induction, we have successfully eliminated the
requirement for lemmas in existing systems (e.g. \cite{nguyen10shape3,qiu13dryad}) 
for proving the functional correctness
of the programs in Table~\ref{table:academic} and \ref{table:real}.
As already stated in Sec.~\ref{sec:intro}, 
existing systems require lemmas in two common scenarios.
First, it is when the traversal order of the data structures is different
from what suggested by the recursive definitions, e.g. ${\tt OpenBSD/queue.h}$.
Second, it is due to the boundaries caused by
iterative loops or multiple function calls.
One example is $\tt{append}$ function in $\tt{glib/gslist.c}$,
where (in addition to the list definition) the list segment, 
\lseg({\sf head},{\sf last}), is necessary to say 
about the function invariant --- the last node of a non-empty input list
is always reachable from the list's head.
Other examples are to make a connection between a sorted list 
and a singly-linked list (e.g. in sorting algorithms), between
two sorted partitions (e.g. in {\tt quick\_sort\_iter}),
between a circular list and a list segment (e.g. {\tt count}),
etc.


\begin{table}[tbh]

\begin{center}
\begin{small}
\begin{tabular}{|l|l|c|}
\hline
\textsf{Program} & \textsf{Function}  
& \textsf{T/F}\\
\hline
\hline

\begin{tabular}{@{}l@{}}\stuff{\bf{glib/gslist.c}}\\ \stuff{Singly Linked-List}\end{tabular}
&
\stuff{find, position, index,\\ 
nth,last,length,append,\\
insert\_at\_pos,merge\_sort,\\
remove,insert\_sorted\_list}
&
${<} 1s$ \\

\hline

\begin{tabular}{@{}l@{}}\stuff{\bf{glib/glist.c}}\\ \stuff{Doubly Linked-List}\end{tabular}
&
\stuff{nth, 
position, find, \\index, last, length}
&
${<} 1s$ \\

\hline

\begin{tabular}{@{}l@{}}\stuff{\bf{OpenBSD/queue.h}}\\ \stuff{Queue}\end{tabular}
&
\stuff{simpleq\_remove\_after, \\simpleq\_insert\_tail, \\
simpleq\_insert\_after}
&
${<} 1s$ \\

\hline

\stuff{\bf{ExpressOS/cachePage.c}}
& \stuff{lookup\_prev,add\_cachepage}
& ${<} 1s$ \\

\hline



\stuff{\bf{linux/mmap.c}}
& \stuff{insert\_vm\_struct}
& ${<} 1s$ \\



\hline
\end{tabular}
\end{small}
\end{center}
\vspace{-8pt}
\caption{Verification of Open-Source Libraries (existing systems require lemmas).}
\label{table:real}
\end{table}

\noindent \emph{\underline{Remark \#2:}} ~The verification time for each function 
is always less than 1 second. This is within our expectation since
when our proof method succeeds, the size of the proof tree is relatively small.
For example, in order to prove the functional correctness of 
$\tt{append}$ function in $\tt{glib/gslist.c}$,
we only need to prove 3 obligations, each of which requires no more than
two left unfolds, two right unfolds and two inductions\footnote{Since 
the number of rules (disjuncts) in a predicate definition is fixed and small,
the size of proof tree mainly depends on the number of unfolds and inductions.}.
In fact, the maximum number of left unfolds, right unfolds 
and inductions used in our system are 5, 5 and 3 respectively,
even for the functions that take U+M frameworks much longer time to prove.
For example, consider $\tt{simpleq\_insert\_after}$, 
a function to insert an element into a queue.
This example requires reasoning about unmatchable predicates: to prove it
\dryad{} needs 18 seconds and the help from a lemma.
Such inefficiency is due to the use of a complicated lemma\footnote{We believe
that the lemmas in \cite{qiu13dryad} are unnecessarily too complicated, 
because the authors want to reduce the number of them, by grouping a few into one.},
which consists of a large disjunction. 
Though efficient in practice, \smt{} solvers
still face a combinatorial explosion challenge as they dissect the disjunction.
In other words, in addition to having a higher level of automation,
our framework has a potential advantage of being more efficient 
than existing U+M systems.


\section{Related Work}
\label{sec:related}
There is a vast literature on program verification considering data structures.
The well known formalism of Separation Logic (SL)~\cite{reynolds02separation} is often combined 
with a recursive formulation of data structure properties.
Implementations, however, are incomplete, e.g., \cite{berdine05symexec,iosif13treewidth}, or deal only with
fragments \cite{berdine03decidable,magill08shape}.
There is also literature on decision procedures for restricted heap
logics; we mention just a few examples:
\cite{rakamaric07heaps,rakamaric07heapsSMT,lahiri08future,ranise06list,bouajjani09ds,bjorner09fp}.
These have, however, severe restrictions on expressivity.
None of them can handle the VC's of the kind considered in this paper.

There is also a variety of verification tools based on classical
logics and \smt{} solvers.  Some examples are 
Dafny \cite{leino10dafny}, 
VCC \cite{cohen09vcc} and
Verifast \cite{jacobs11verifast} 
which require significant ghost
annotations, and annotations that explicitly express and manipulate
frames.  They do not automatically verify the general and 
complex properties addressed in this paper, but in general
resort to interactive theorem provers, e.g. Mona, Isabelle or Coq,
which usually requires manual guidance.

In \cite{navarro11pldi}, Navarro and Rybalchenko showed that significant performance improvements can be 
obtained by incorporating first-order theorem proving techniques into SL provers. 
However, the focus of that work is about list segments, not general user-defined recursive predicates.  
On a similar thread, \cite{wies13cav} advances the automation of SL, using \smt{}, 
in verifying procedures manipulating list-like data structures.
The works \cite{zee08linked,zee09imperative,nguyen10shape3,madhusudan12dryad,qiu13dryad} 
are also close related works: they form the U+M paradigm which we have
carefully discussed in Section~\ref{sec:intro} and~\ref{sec:example}.

We have discussed the works on automatic (and explicit)
induction~\cite{ACL1,ACL2,leino12vmcai,ZENO} in Section~\ref{sec:example}. 
Here we further highlight the work of Lahiri and Qadeer~\cite{lahiri06popl}, 
which adapts the induction principle for proving properties of well-founded linked list.
The technique relies on the well-foundedness of the heap,
while employing the induction principle to derive from two basic axioms 
a small set of additional first-order axioms that are useful for 
proving the correctness of several simple programs.

\ignore{

Our contribution is mainly based on the automatic use of induction,
thus breaking through the U+M barrier.
In the literature, there have been works on automatic induction~\cite{ACL1,ACL2,leino12vmcai,ZENO}. 
They are concerned with
proving a \emph{fixed} hypothesis, say $h(\tilde{x})$, that is, to show
that $h()$ holds over all values of the variables $\tilde{x}$.
The challenge is to discover and prove
$h(f(\tilde{x}))~{\implies}~h(g(\tilde{x}))$, where $f(\tilde{x})$ and $g(\tilde{x})$
are expressions involving $\tilde{x}$ and the former expression is
less than the latter in some well-founded measure.
Furthermore, a \emph{base case} of $h()$ needs to be proven.
In contrast, our notion of induction hypothesis is completely different.
Our hypothesis is discovered \emph{dynamically}, that is, it
can be any formula arising from the proof search process.
}

Closer to the spirit of our work, there are works on 
``Cyclic Proof''~\cite{cyclic11cade,cyclic12aplas} (which we have discussed) 
and ``Matching Logic''~\cite{matching-logic}.
They are based on the same principle that when two similar obligations are
detected in the same proof path, the latter can be used to subsume to former.
To some extent, these are special and limited instances of our induction rules.
In terms of implementation, the ``circularity rule'' in~\cite{matching-logic}
can only be applied to basic ``patterns'' in the logic, therefore it
cannot support general user-defined recursive predicates. 

We finally mention the work \cite{jaffar08coind}, from which the concept of 
our automatic induction originates. 
The current paper extends \cite{jaffar08coind} first by refining the original single
coinduction rule into two more powerful rules, to deal with the antecedent and
consequent of a VC respectively.
Secondly, the \emph{application} of the rules has been systematized so as to
produce a rigorous proof search strategy.
Another technical advance is our introduction of \emph{timestamps} 
(a progressive measure) in the two induction rules 
as an efficient technique to avoid circular reasoning. 
Finally, the present paper focuses on program verification
and uses a specific domain of discourse involving the use 
of explicit symbolic heaps and separation.

\section{Concluding Remarks}
\label{sec:conclusion}
We presented a framework for proving recursive properties 
of data structures.
The main contribution was an algorithm which
provided a new level of automation across a wider class of programs.
Its key technical features were two automatic re-writing rules,
based on a systematic consideration of dynamically generated
possibilities as induction hypotheses.  Finally, experimental evidence
showed that the algorithm has gone beyond
the state-of-the-art.  



\balance

\bibliographystyle{plain}
\bibliography{references}

%




\end{document}